 \documentclass[final,3p,times,number,sort,compress]{elsarticle}

\usepackage{multirow}
\usepackage{amssymb}
\usepackage{lipsum}
\usepackage{amsmath}
\usepackage{booktabs}
\usepackage{float}
\usepackage{graphicx}
\usepackage{subcaption}
\usepackage{upgreek}
\usepackage{placeins}
\usepackage{algorithm}
\usepackage{algpseudocode}
\usepackage{makecell}
\usepackage[font=normalsize, labelfont=bf]{caption}
\usepackage[font=normalsize]{subcaption}
\usepackage[labelfont=normalfont]{subcaption}

\usepackage{xcolor}

\journal{Elsevier}

\begin{document}

\begin{frontmatter}



\title{Accelerating phase-field simulations of grain growth via tensor decomposition}


\author[first]{Chaoqian Yuan}
\author[first]{Dingxiang Zhu}
\author[first]{Ye Lu\corref{cor1}}
\ead{yelu@umbc.edu}
\affiliation[first]{organization={Department of Mechanical Engineering, University of Maryland Baltimore County},
            city={Baltimore},
            country={USA}}
 \cortext[cor1]{Corresponding author}

\begin{abstract}
Phase-field models provide a powerful framework for simulating grain growth phenomena in polycrystalline materials, but their application to large-scale systems is limited by the high computational and memory costs associated with resolving multiple phase variables over fine spatial and temporal scales. This paper presents a tensor decomposition based model reduction framework for accelerating phase-field simulations of grain growth. The proposed method exploits the underlying low-rank structure of the evolving phase-field variables and represents the high-dimensional solution in a spatially decomposed tensor format. By performing the numerical computations directly within this reduced representation, the framework decreases storage requirements and reduces the cost of repeatedly solving the phase-field equations. The number of modes for the decomposed approximation is controlled during the online computation with a subspace mode-updating algorithm, to balance computational efficiency and approximation accuracy. The method is assessed using representative grain growth problems, and its predictions are compared with those obtained from conventional full-order finite element simulations. The results demonstrate that the formulation can accurately reproduce the evolution of grain morphology and the principal characteristics of grain growth kinetics while substantially improving computational efficiency. The proposed framework therefore provides a promising approach for extending phase-field simulations to larger spatial domains, finer discretizations, and systems containing a large number of phase variables.
\end{abstract}


\end{frontmatter}


\section{Introduction}
\label{introduction}

The microstructure of materials plays a critical role in determining their mechanical, thermal, and functional properties. Understanding and predicting the formation and evolution of microstructure are therefore crucial for controlling material properties  and guiding the design of manufacturing processes. Grain growth is a fundamental microstructure evolution process in polycrystalline materials, characterized by the growth of some grains at the expense of others, leading to an increase in the average grain size \cite{Rohrer2023,Marder2026,Mullins2001}. This process is primarily driven by the reduction of total grain boundary energy through boundary migration, accompanied by complex morphological and topological changes, including grain shrinkage, disappearance, and junction migration. Given the complexity of the underlying interfacial dynamics, phase-field methods provide an attractive computational framework to simulate grain growth, as grain boundaries are implicitly represented by continuous phase-field variables (namely order parameters), avoiding the challenges associated with explicit interface tracking \cite{chen2002phase,boettinger2002phase,Moelans2008}. Furthermore, phase-field methods can readily accommodate coupling among thermal, chemical, mechanical, and electrical fields, enabling the simulation of complex multiphysics phenomena. This versatility has led to their widespread applications in modeling microstructure evolution across various materials and manufacturing processes, including solidification in additive manufacturing \cite{Chadwick2021,  chadwick2025microstructure,yuan2026efficient} and solid-state sintering \cite{wang2006computer,biswas2018phase,hotzer2019phase}.

Despite these advantages, large-scale  phase-field simulations remain computationally expensive and  resource-demanding. In particular, fine spatial discretization is typically required in phase-field models to adequately resolve diffuse interfaces and capture interactions with small-scale particles. Moreover, grain growth models usually involve the simultaneous evolution of numerous order-parameter fields over a large number of time steps. These challenges become particularly pronounced in three-dimensional simulations, where computational cost and memory requirements increase rapidly with spatial resolution and the number of grains \cite{Flint2018,Yang2021,Kim2006,Dimokrati2020}. Consequently, large-scale phase-field simulations of grain growth often require substantial high-performance computing infrastructure \cite{Miyoshi2017,Suwa2008,Miyoshi2021Aniso,Sitompul2022}, thereby limiting their practical applicability to extensive parametric studies, iterative design, and process optimization.

Various numerical techniques have been developed to improve the efficiency of phase-field simulations. These include adaptive meshing \cite{provatas1998efficient,zhu2019phase,dewitt2020prisms}, higher-order spatial approximation \cite{feng2006spectral,li2024adaptive}, and efficient time integration algorithms \cite{graser2013time,shen2010numerical,zhao2017numerical}. Specifically, adaptive meshing reduces the number of spatial degrees of freedom (DoFs) by concentrating computational resources in regions requiring higher resolution, while higher-order approximations can improve spatial accuracy with relatively fewer degrees of freedom. Meanwhile, advanced time integration schemes can improve computational efficiency by allowing larger time steps while maintaining numerical stability and accuracy. Despite these advances, the computational complexity associated with resolving evolving grain boundaries and simultaneously updating numerous coupled order-parameter fields remains a significant challenge, particularly for large-scale three-dimensional simulations. Furthermore, conventional numerical methods can still incur substantial memory and data storage requirements, which increase rapidly with the number of spatial DoFs, order-parameters, and time steps.

More recently, machine learning techniques have also been explored to accelerate phase-field simulations \cite{wight2020solving,liu2019multi,montes2021accelerating,peivaste2022machine,hu2022accelerating,oommen2022learning,ciesielski2025deep,peivaste2025teaching}. Many of these approaches employ surrogate models trained on precomputed simulation data to predict microstructure evolution at a substantially reduced computational cost. However, their predictive accuracy and generalization capabilities strongly depend  on the representativeness of the training data, potentially limiting their applicability when problem settings, such as material parameters, initial microstructures, and multiphysics coupling, differ significantly from those encountered during training. Moreover, generating sufficiently comprehensive training datasets through high-fidelity phase-field simulations and training neural networks can still be computationally demanding. Although physics-informed machine learning approaches \cite{raissi2019physics,mattey2022novel,li2023phase} can reduce the dependence on precomputed simulation data by incorporating governing equations and physical constraints into the learning process, their computational efficiency and robustness for large-scale, three-dimensional microstructure evolution remain important challenges.

Model reduction techniques offer a promising alternative to reduce computational cost for large-scale problems \cite{maday2002reduced,willcox2002balanced,quarteroni2011certified,benner2017model,choi2021space,zhang2022hidenn,li2023convolution}. In the context of phase-field simulations, relevant work include the development of proper orthogonal decomposition method \cite{song2016reduced,uzunca2017energy,zhou2019reduced,li2021numerical,wu2023certified} and the generalized multiscale finite element method \cite{tyrylgin2021multiscale}. Despite their great potential, these approaches typically require an offline stage involving the generation of solution snapshots and an appropriate reduced basis, which can still be computationally expensive, particularly when high-fidelity phase-field simulations are needed. Such offline computational requirements may limit their efficiency and flexibility across different problem settings, similar to conventional machine learning approaches. These limitations motivate the development of alternative computational frameworks that eliminate the need for precomputed high-fidelity simulation data while substantially reducing computational costs and memory requirements without compromising the accuracy of physics-based predictions.

In this work, we explore a tensor decomposition based model reduction framework for accelerating large-scale phase-field simulations. Building upon our previous development for the general nonlinear Allen--Cahn equation \cite{Lu2025}, we extended the framework to multiphase-field simulations of grain growth involving numerous coupled order-parameter fields and complex physics-based energy functionals. By exploiting the underlying low-rank structure of phase-field solutions through a separation of variables, the proposed approach enables compact representations of evolving grain structures, substantially reducing the computational cost and data size associated with conventional numerical methods. Unlike conventional data-driven approaches, the framework directly solves the governing phase-field equations without requiring precomputed high-fidelity simulation data or offline training. Furthermore, we develop a general subspace mode-updating algorithm to reduce the number of modes required for accurate low-rank approximations. The detailed formulation and GPU implementation of the proposed framework will be presented, followed by representative grain growth simulations to evaluate its computational efficiency and accuracy against reference finite element method (FEM). The main contributions of this work are summarized as follows:
\begin{itemize}
    \item A tensor decomposition based model reduction framework for efficient large-scale multiphase-field simulations involving numerous coupled order-parameter fields.
    \item A general subspace mode-updating algorithm to reduce the number of modes required for accurate low-rank approximations.
    \item GPU-accelerated simulations of competitive grain growth in polycrystalline materials, demonstrating the computational efficiency and accuracy of the proposed framework through comparisons with conventional FE simulations.
\end{itemize}

The remainder of this paper is organized as follows. Section 2 introduces a representative phase-field model for grain growth and microstructure evolution, along with its FE formulation. Section 3 presents the proposed tensor decomposition based computational framework, including the detailed solution algorithm and GPU implementation. Section 4 presents three-dimensional grain growth simulations to evaluate the accuracy and computational efficiency of the proposed approach. Finally, Section 5 summarizes the main findings and provides concluding remarks.

\section{Problem formulation}
\subsection{Phase field model of grain growth}
The general phase-field model of grain growth can be written as a multiphase Allen--Cahn type equation, which reads
\begin{equation}
\label{eq:2.1_allen_cahn}
\frac{\partial \phi_i}{\partial t} + L\left(\frac{\partial f\left(\{\phi\}\right)}{\partial \phi_i} - \kappa\Delta \phi_i \right)=0, 
\end{equation}
where $\phi_i$ denotes the order parameters (OPs) representing grains of different crystal orientations, $\{\phi\}=\{\phi_1,\phi_2,\ldots,\phi_N\}$ denotes the set of all OPs, $L$ is the mobility, $f\left(\{\phi\}\right)$ is a homogeneous free energy, and $\kappa$ is the gradient-energy coefficient. By  definition, the evolution of each OP is equivalent to a gradient flow problem \cite{chen2002phase,boettinger2002phase} and is driven by the reduction of a total free energy functional $E$ that usually consists of bulk and gradient energy contributions. In this work, we adopt a commonly used free energy definition for $E$ (e.g., \cite{chadwick2025microstructure,yuan2026efficient})

\begin{equation}
\label{eq:2.1_total_free_energy}
E = \int_{\Omega} \left[ f(\{\phi\}) + \frac{\kappa}{2} \sum_{i=1}^{N}|\nabla \phi_i|^2\right] ~ d\Omega,
\end{equation}
with the homogeneous energy $f\left(\{\phi\}\right)$ defined as

\begin{equation}
\label{eq:2.1_homogeneous_free_energy}
f(\{\phi\}) = W\left[\sum_{i=1}^{N}\left(\frac{\phi_i^4}{4}-\frac{\phi_i^2}{2} \right) +\gamma_{PF}\sum_{i=1}^{N}\sum_{j>i}^{N}\phi_i^2\phi_j^2 +\frac{1}{4} \right],
\end{equation}
where $W$ is the height of the entire multi-well energy, and $\gamma_{PF}$ is a penalty coefficient. The first term is known as a double-well potential that stabilizes the bulk phases by driving each OP toward its equilibrium values, and the second term is preventing the overlap of different OPs. This definition makes a highly nonlinear and transient problem that requires appropriate spatial discretizations and time integration algorithms that guarantee the physically-consistent energy reduction with
\begin{equation}
\label{eq:energy_law}
E(t_{k+1})-E(t_{k})\leq0,
\end{equation}
where $t_{k}$ and $t_{k+1}$ denote two consecutive time steps. Various semi-implicit or implicit time integration schemes \cite{graser2013time,shen2010numerical,zhao2017numerical} can employed to ensure this discrete energy law. In the following, we present a FEM coupled with an unconditionally stable semi-implicit scheme \cite{Lu2025}.


\subsection{Finite element formulation}
Now, let us consider a weak formulation of the phase-field model.  By multiplying Eq.~\eqref{eq:2.1_allen_cahn} with a test function $\delta\phi_i$ and integrating over the computational domain $\Omega$, the weak form of the problem can be obtained as follows
\begin{equation}
\label{eq:2.2_weak_form}
\int_\Omega \delta \phi_i\frac{d \phi_i}{d t} ~ d\Omega + \int_\Omega \delta \phi_i ~ L\frac{\partial f\left(\{\phi\}\right)}{\partial \phi_i} ~ d\Omega + \int_\Omega  L\ \kappa\ \nabla \delta \phi_i \cdot  \nabla \phi_i ~ d\Omega = 0, \quad \forall t \in [0, t_f],
\end{equation}
with the natural boundary condition $\nabla\phi_i\cdot n=0$ applied on $\partial\Omega$.  The spatial discretization can then be performed using the FE approximation of the OP fields

\begin{equation}
\label{eq:2.3_fe_approximation}
\phi_{i}(x,y,z) = \boldsymbol{N} (x,y,z)~\boldsymbol{\phi}_i,
\end{equation}
where $(x,y,z)$ denotes the spatial coordinate,  $\boldsymbol{N}$ is the FE shape function, and $\boldsymbol{\phi}_i$ is the nodal solution vector of the $i$-th OP. With the same FE approximation for the test function $\delta \phi=\boldsymbol{N}\delta\boldsymbol{\phi}_i$, Eq.~\eqref{eq:2.2_weak_form} becomes

\begin{equation}
\label{eq:2.3_discrete_weak_form}
\int_\Omega \delta \boldsymbol{\phi}_i^T \boldsymbol{N}^T\frac{d {\phi}_i}{d t} ~ d\Omega + \int_\Omega \delta \boldsymbol{\phi}_i^T \boldsymbol{N}^T ~ L \frac{ \partial f\left(\{\boldsymbol{\phi}\}\right)}{\partial {\phi}_i} ~ d\Omega + \int_\Omega \delta \boldsymbol{\phi}_i^T \boldsymbol{B}^T L \ \kappa\ \boldsymbol{B} \boldsymbol{\phi}_i ~ d\Omega = 0, \quad \forall t \in [0, t_f],
\end{equation}
where $\boldsymbol{B}$ is the gradient of the shape function. Therefore, the semi-discrete form of the problem is obtained as 
\begin{equation}
\label{eq:FEsemi}
\int_{\Omega}\boldsymbol{N}^T \frac{\partial \phi_i}{\partial t} \, d\Omega + \boldsymbol{Q}_i ({\{\boldsymbol{\phi}\}}) + \boldsymbol{K}\ \boldsymbol{\phi}_i =0, \quad \forall t \in [0, t_f],
\end{equation}
with 
\begin{equation}
\left\{
\begin{aligned}
     & \boldsymbol{Q}_i ({\{\boldsymbol{\phi}\}})=\int_{\Omega}\boldsymbol{N}^T L  f'(\{\boldsymbol{\phi}\})\, d\Omega,\\
     &\boldsymbol{K}=  \int_{\Omega}\boldsymbol{B}^T L \ \kappa \boldsymbol{B}\, d\Omega,
\end{aligned}
\right.
\end{equation}
where $f'(\{\boldsymbol{\phi}\})=\frac{ \partial f\left(\{\boldsymbol{\phi}\}\right)}{\partial {\phi}_i}$. An implicit time integration could be used to solve the above equation. However, repetitively updating the nonlinear term in the equation at each time step can be extremely expensive, especially when considering that the evolution of grains may involve thousands of time increments.

\subsection{Stabilized semi-implicit time integration}
Following our previous work \cite{Lu2025}, we can adopt a stabilized semi-implicit scheme for the time integration, which can  unconditionally ensure the discrete energy law \eqref{eq:energy_law} and can be easily implemented without introducing auxiliary variables. The discretized formulation is written as

\begin{equation}
\label{eq:3.2_stabilized_semi_implicit}
\int_\Omega \boldsymbol{N}^T\boldsymbol{N} (\boldsymbol{\phi}^{k+1}_i-\boldsymbol{\phi}^{k}_i)(\frac{1}{\Delta t}+\alpha L) ~ d\Omega + \boldsymbol{Q}_i ({\{\boldsymbol{\phi}^{k}\}}) + \boldsymbol{K}\ \boldsymbol{\phi}^{k+1}_i =0,
\end{equation}
where $\alpha$ is a stabilizing parameter, $\boldsymbol{\phi}^{k+1}_i$ is the solution at the current time $t_{k+1}$, and $\boldsymbol{\phi}^{k}_i$ is the solution at the previous step $t_k$. The final discrete equation is then
\begin{equation}
\label{eq:3.2_stabilized_semi_implicit_discrete}
\boldsymbol{M} \boldsymbol{\phi}_i^{k+1}- \boldsymbol{M}\boldsymbol{\phi}_i^{k} 
+  \boldsymbol{Q}_i (\{\boldsymbol{\phi}^k\}) + \boldsymbol{K}\ \boldsymbol{\phi}^{k+1}_i =0, 
\end{equation}
with 
\begin{equation}
\boldsymbol{M} =\int_{\Omega}\boldsymbol{N}^T \boldsymbol{N} (\frac{1}{\Delta t} + \alpha L) \, d\Omega. 
\end{equation}

The stabilizing parameter $\alpha$ can be chosen as 
\begin{equation}
\label{eq:3.2_stabilization_condition}
\alpha \geq \frac{1}{2}\sup\left(\frac{\partial^2 f\left(\{\phi\}\right)}{\partial \phi_i^2}\right), \qquad \forall \phi_i\in[0,1],
\end{equation}
which can theoretically guarantee the discrete energy law for an arbitrary step size $\Delta t$.  A  proof can be found in our previous work \cite{Lu2025}. As mentioned earlier, this problem can still require substantial computational resources, due to the fine meshes needed to resolve the length scale of grain evolution and a large number of OPs in three-dimensional simulations.

\section{The tensor decomposition based model reduction for grain growth}
\subsection{Spatially decomposed low-rank approximation}
To reduce the computational cost with large-scale phase-field simulations, we can develop the tensor decomposition (TD) based model reduction technique using a spatially decomposed representation of OPs
\begin{equation}
\label{eq:3.1_spatial_decomposition}
\phi_i(x,y,z) \approx \phi^{\text{TD}}= \sum_{m=1}^M u^{(m)}(x)\ v^{(m)}(y)\ w^{(m)}(z),
\end{equation}
where $\phi_i$ is the solution at any given time step,  $u^{(m)}$, $v^{(m)}$, and $w^{(m)}$ are the separated one-dimensional functions associated with the $x$-, $y$-, and $z$-directions respectively, and $M$ is the total number of modes. This separation of variables corresponds to the canonical polyadic decomposition of tensors \cite{Kolda2009} and can be determined through either data-driven machine learning approaches \cite{lu2018adaptive,lu2018multi,lu2019datadriven} or data-free manners by solving the governing equations \cite{Chinesta2011,Chinesta2017,lu2023convolution,Lu2024}. In this work, we develop the detailed  formulation to determine this spatially decomposed low-rank approximation by solving the governing equation of grain growth. The final solution of the problem becomes finding the separated one-dimensional functions  $u^{(m)}$, $v^{(m)}$, and $w^{(m)}$, rather than the original solution $\phi_i$. It is clear that if $M$ increases, the solution of TD, i.e., $\phi^{\text{TD}}$, converges to $\phi_i(x,y,z)$, even if the original solution cannot be analytically written as products of separated functions. Some theoretical proofs  can be found in \cite{zhang2022hidenn}.

 This spatial decomposition  offers several computational advantages over conventional numerical methods. First, it transforms the solution of a 3D problem into a sequence of 1D problems, leading to substantial reductions in both computational cost and memory consumption. For large-scale simulations, even when iterative solvers are employed, the memory required to store the linear system and the associated solver data often becomes a major bottleneck, increasing rapidly with the number of DoFs. In contrast, the memory required by the TD method grows relatively slowly and only scales with the size of those discretized 1D functions. Second, a deep reduction in solution data size can be expected. Since transient simulations require the solution to be transferred from one time step to another, storing only the 1D functions greatly reduces the amount of data that must be retained and communicated. This advantage becomes even more pronounced in distributed-memory parallel computing, such as MPI-based implementations, where the communication cost associated with inter-node data transfer can become an important factor in overall computational performance.

\subsection{Model reduction for grain growth}
Based on the spatial decomposition of the OPs, we can develop a novel discrete formulation to compute each mode of the 1D function functions $u^{(m)}$, $v^{(m)}$, and $w^{(m)}$ by solving the  governing equation for grain growth.  To simplify the notation, we can first rewrite Eq.~\eqref{eq:3.1_spatial_decomposition} in an incremental form by assuming the $M-1$ modes are already computed

\begin{equation}
\label{eq:3.3_mode_enrichment}
\phi^{\text{TD}} = \sum_{m=1}^{M-1} u^{(m)} v^{(m)} w^{(m)} + u\ v\ w,
\end{equation}
where $u$, $v$, and $w$ are the solutions for the $M$-th mode. The test function in Eq.~\eqref{eq:2.2_weak_form} in this case becomes
\begin{equation}
\label{eq:3.3_test_function_variation}
\delta\phi = \delta u\ v\ w + u \ \delta v \ w + u\ v \ \delta w. 
\end{equation}

Substituting Eqs.~\eqref{eq:3.3_mode_enrichment} and \eqref{eq:3.3_test_function_variation} into  the weak formulation of the problem leads to the solutions for $u$, $v$, and $w$. For the  stabilized semi-implicit formulation given below

\begin{equation}
\label{eq:3.3_td_weak_form}
\int_\Omega \delta\phi (\phi^{k+1}_i-\phi^{k}_i)(\frac{1}{\Delta t}+\alpha L) ~ d\Omega + \int_\Omega \delta\phi ~ L f'(\{\phi^{k}\}) ~ d\Omega + \int_\Omega \nabla\delta\phi \cdot L\kappa \nabla \phi^{k+1}_i ~ d\Omega = 0,
\end{equation}
we can consider $\phi_i^{k+1} = \phi^{\text{TD}} = \sum_{m=1}^{M-1} u^{(m)} v^{(m)} w^{(m)} + u\ v\ w$, and $\phi_i^{k}$ is already known as $\phi_i^{k} = \sum_{m=1}^{M_k} u^{(m,k)} v^{(m,k)} w^{(m,k)}$ at the current time step. Following the derivation in \cite{Lu2025} and using the 1D FE approximation for $u$, $v$, and $w$ 
\begin{equation}
    {u}=\boldsymbol{N}_x\ \boldsymbol{u},\qquad
    {v}=\boldsymbol{N}_y\ \boldsymbol{v},\qquad
    {w}=\boldsymbol{N}_z\ \boldsymbol{w},
\end{equation}
we can obtain three discrete systems of equations for the nodal solutions $\boldsymbol{u}$, $\boldsymbol{v}$, and $\boldsymbol{w}$
\begin{equation}
\label{eq:3.3_directional_systems}
\boldsymbol{K}_u\ \boldsymbol{u} = \boldsymbol{Q}_u, \qquad \boldsymbol{K}_v\ \boldsymbol{v} = \boldsymbol{Q}_v,\qquad \boldsymbol{K}_w \ \boldsymbol{w} = \boldsymbol{Q}_w,
\end{equation}
with
\begin{equation}
\label{eq:3.3_u_direction_system}
\left\{
\begin{aligned}
\boldsymbol{K}_u
&= \left(\frac{1}{\Delta t}+\alpha L\right)
\boldsymbol{M}_{xx}\ \boldsymbol{v}^T\boldsymbol{M}_{yy}\boldsymbol{v}
\ \boldsymbol{w}^T\boldsymbol{M}_{zz}\boldsymbol{w}
\\
&\quad + L\kappa\
\boldsymbol{K}_{xx}\ \boldsymbol{v}^T\boldsymbol{M}_{yy}\boldsymbol{v}\ 
\boldsymbol{w}^T\boldsymbol{M}_{zz}\boldsymbol{w}\\
&\quad + L\kappa\
\boldsymbol{M}_{xx}\ \boldsymbol{v}^T\boldsymbol{K}_{yy}\boldsymbol{v}\
\boldsymbol{w}^T\boldsymbol{M}_{zz}\boldsymbol{w}\\
&\quad + L\kappa\
\boldsymbol{M}_{xx}\ \boldsymbol{v}^T\boldsymbol{M}_{yy}\boldsymbol{v}\
\boldsymbol{w}^T\boldsymbol{K}_{zz}\boldsymbol{w}
,
\\[0.5em]
\boldsymbol{Q}_u
&=
\left(\frac{1}{\Delta t}+\alpha L\right)
\sum_{m=1}^{M_k}
\boldsymbol{M}_{xx}\boldsymbol{u}^{(m,k)}\ 
\boldsymbol{v}^T\boldsymbol{M}_{yy}\boldsymbol{v}^{(m,k)}\ 
\boldsymbol{w}^T\boldsymbol{M}_{zz}\boldsymbol{w}^{(m,k)}
\\
&\quad
-\left(\frac{1}{\Delta t}+\alpha L\right)
\sum_{m=1}^{M-1}
\boldsymbol{M}_{xx}\boldsymbol{u}^{(m)}\ 
\boldsymbol{v}^T\boldsymbol{M}_{yy}\boldsymbol{v}^{(m)}\ 
\boldsymbol{w}^T\boldsymbol{M}_{zz}\boldsymbol{w}^{(m)}
\\
&\quad
-L\sum_{m=1}^{M_f} \boldsymbol{M}_{xx} \boldsymbol{q}^{(m)}_x \ \boldsymbol{v}^T\boldsymbol{M}_{yy} \boldsymbol{q}^{(m)}_y \  \boldsymbol{w} ^T\boldsymbol{M}_{zz} \boldsymbol{q}^{(m)}_z
\\
&\quad
-L\kappa
\sum_{m=1}^{M-1}
\boldsymbol{K}_{xx}\boldsymbol{u}^{(m)}\
\boldsymbol{v}^T\boldsymbol{M}_{yy}\boldsymbol{v}^{(m)}\
\boldsymbol{w}^T\boldsymbol{M}_{zz}\boldsymbol{w}^{(m)}
\\
&\quad
-L\kappa
\sum_{m=1}^{M-1}
\boldsymbol{M}_{xx}\boldsymbol{u}^{(m)}\
\boldsymbol{v}^T\boldsymbol{K}_{yy}\boldsymbol{v}^{(m)}\
\boldsymbol{w}^T\boldsymbol{M}_{zz}\boldsymbol{w}^{(m)}
\\
&\quad
-L\kappa
\sum_{m=1}^{M-1}\boldsymbol{M}_{xx}\boldsymbol{u}^{(m)}\
\boldsymbol{v}^T\boldsymbol{M}_{yy}\boldsymbol{v}^{(m)}\
\boldsymbol{w}^T\boldsymbol{K}_{zz}\boldsymbol{w}^{(m)},
\end{aligned}
\right.
\end{equation}
\begin{equation}
\label{eq:3.3_v_direction_system}
\left\{
\begin{aligned}
\boldsymbol{K}_v
&= \left(\frac{1}{\Delta t}+\alpha L\right)
\boldsymbol{M}_{yy} \ \boldsymbol{u}^T\boldsymbol{M}_{xx}\boldsymbol{u}\
\boldsymbol{w}^T\boldsymbol{M}_{zz}\boldsymbol{w} \\
&\quad + L\kappa\
\boldsymbol{K}_{yy}\ \boldsymbol{u}^T\boldsymbol{M}_{xx}\boldsymbol{u}\
\boldsymbol{w}^T\boldsymbol{M}_{zz}\boldsymbol{w}\\
&\quad + L\kappa\
\boldsymbol{M}_{yy}\ \boldsymbol{u}^T\boldsymbol{K}_{xx}\boldsymbol{u}\
\boldsymbol{w}^T\boldsymbol{M}_{zz}\boldsymbol{w}\\
&\quad + L\kappa\
\boldsymbol{M}_{yy}\ \boldsymbol{u}^T\boldsymbol{M}_{xx}\boldsymbol{u}\
\boldsymbol{w}^T\boldsymbol{K}_{zz}\boldsymbol{w}
,
\\[0.5em]
\boldsymbol{Q}_v
&=
\left(\frac{1}{\Delta t}+\alpha L\right)
\sum_{m=1}^{M_k}
\boldsymbol{M}_{yy}\boldsymbol{v}^{(m,k)}
\boldsymbol{u}^T\boldsymbol{M}_{xx}\boldsymbol{u}^{(m,k)}
\boldsymbol{w}^T\boldsymbol{M}_{zz}\boldsymbol{w}^{(m,k)}
\\
&\quad
-\left(\frac{1}{\Delta t}+\alpha L\right)
\sum_{m=1}^{M-1}
\boldsymbol{M}_{yy}\boldsymbol{v}^{(m)}\
\boldsymbol{u}^T\boldsymbol{M}_{xx}\boldsymbol{u}^{(m)}\
\boldsymbol{w}^T\boldsymbol{M}_{zz}\boldsymbol{w}^{(m)}
\\
&\quad
-L\sum_{m=1}^{M_f} \boldsymbol{M}_{yy} \boldsymbol{q}^{(m)}_y \ \boldsymbol{u}^T\boldsymbol{M}_{xx} \boldsymbol{q}^{(m)}_x \  \boldsymbol{w} ^T\boldsymbol{M}_{zz} \boldsymbol{q}^{(m)}_z
\\
&\quad
-L\kappa
\sum_{m=1}^{M-1}
\boldsymbol{K}_{yy}\boldsymbol{v}^{(m)}\
\boldsymbol{u}^T\boldsymbol{M}_{xx}\boldsymbol{u}^{(m)}\
\boldsymbol{w}^T\boldsymbol{M}_{zz}\boldsymbol{w}^{(m)}
\\
&\quad
-L\kappa
\sum_{m=1}^{M-1}
\boldsymbol{M}_{yy}\boldsymbol{v}^{(m)}\
\boldsymbol{u}^T\boldsymbol{K}_{xx}\boldsymbol{u}^{(m)}\
\boldsymbol{w}^T\boldsymbol{M}_{zz}\boldsymbol{w}^{(m)}
\\
&\quad
-L\kappa
\sum_{m=1}^{M-1}
\boldsymbol{M}_{yy}\boldsymbol{v}^{(m)}\
\boldsymbol{u}^T\boldsymbol{M}_{xx}\boldsymbol{u}^{(m)}\
\boldsymbol{w}^T\boldsymbol{K}_{zz}\boldsymbol{w}^{(m)},
\end{aligned}
\right.
\end{equation}
and
\begin{equation}
\label{eq:3.3_w_direction_system}
\left\{
\begin{aligned}
\boldsymbol{K}_w
&=
\left(\frac{1}{\Delta t}+\alpha L\right)
\boldsymbol{M}_{zz}\ \boldsymbol{u}^T\boldsymbol{M}_{xx}\boldsymbol{u}\ 
\boldsymbol{v}^T\boldsymbol{M}_{yy}\boldsymbol{v}
\\
&\quad
+L\kappa\
\boldsymbol{K}_{zz}\ \boldsymbol{u}^T\boldsymbol{M}_{xx}\boldsymbol{u}\ 
\boldsymbol{v}^T\boldsymbol{M}_{yy}\boldsymbol{v}
\\
&\quad
+L\kappa\
\boldsymbol{M}_{zz}\ \boldsymbol{u}^T\boldsymbol{K}_{xx}\boldsymbol{u}\
\boldsymbol{v}^T\boldsymbol{M}_{yy}\boldsymbol{v}
\\
&\quad
+L\kappa\
\boldsymbol{M}_{zz}\ \boldsymbol{u}^T\boldsymbol{M}_{xx}\boldsymbol{u}\
\boldsymbol{v}^T\boldsymbol{K}_{yy}\boldsymbol{v}
,
\\[0.5em]
\boldsymbol{Q}_w
&=
\left(\frac{1}{\Delta t}+\alpha L\right)
\sum_{m=1}^{M_k}
\boldsymbol{M}_{zz}\boldsymbol{w}^{(m,k)}\ 
\boldsymbol{u}^T\boldsymbol{M}_{xx}\boldsymbol{u}^{(m,k)}\ 
\boldsymbol{v}^T\boldsymbol{M}_{yy}\boldsymbol{v}^{(m,k)}
\\
&\quad
-\left(\frac{1}{\Delta t}+\alpha L\right)
\sum_{m=1}^{M-1}
\boldsymbol{M}_{zz}\boldsymbol{w}^{(m)}\ 
\boldsymbol{u}^T\boldsymbol{M}_{xx}\boldsymbol{u}^{(m)}\
\boldsymbol{v}^T\boldsymbol{M}_{yy}\boldsymbol{v}^{(m)}
\\
&\quad
-L\sum_{m=1}^{M_f} \boldsymbol{M}_{zz} \boldsymbol{q}^{(m)}_z \ \boldsymbol{u}^T\boldsymbol{M}_{xx} \boldsymbol{q}^{(m)}_x \ \boldsymbol{v}^T\boldsymbol{M}_{yy} \boldsymbol{q}^{(m)}_y
\\
&\quad
-L\kappa
\sum_{m=1}^{M-1}
\boldsymbol{K}_{zz}\boldsymbol{w}^{(m)}\
\boldsymbol{u}^T\boldsymbol{M}_{xx}\boldsymbol{u}^{(m)}\
\boldsymbol{v}^T\boldsymbol{M}_{yy}\boldsymbol{v}^{(m)}
\\
&\quad
-L\kappa
\sum_{m=1}^{M-1}
\boldsymbol{M}_{zz}\boldsymbol{w}^{(m)}\
\boldsymbol{u}^T\boldsymbol{K}_{xx}\boldsymbol{u}^{(m)}\
\boldsymbol{v}^T\boldsymbol{M}_{yy}\boldsymbol{v}^{(m)}
\\
&\quad
-L\kappa
\sum_{m=1}^{M-1}
\boldsymbol{M}_{zz}\boldsymbol{w}^{(m)}\
\boldsymbol{u}^T\boldsymbol{M}_{xx}\boldsymbol{u}^{(m)}\
\boldsymbol{v}^T\boldsymbol{K}_{yy}\boldsymbol{v}^{(m)},
\end{aligned}
\right.
\end{equation}
where $\boldsymbol{K}_{xx} = \int_{\Omega_x} \boldsymbol{B}_x^T \boldsymbol{B}_x\,dx$, $\boldsymbol{M}_{xx} = \int_{\Omega_x} \boldsymbol{N}_x^T \boldsymbol{N}_x\,dx$, $\boldsymbol{K}_{yy} = \int_{\Omega_y} \boldsymbol{B}_y^T \boldsymbol{B}_y\,dy$, $\boldsymbol{M}_{yy} = \int_{\Omega_y} \boldsymbol{N}_y^T \boldsymbol{N}_y\,dy$, $\boldsymbol{K}_{zz} = \int_{\Omega_z} \boldsymbol{B}_z^T \boldsymbol{B}_z\,dz$, and $\boldsymbol{M}_{zz} = \int_{\Omega_z} \boldsymbol{N}_z^T \boldsymbol{N}_z\,dz$. In addition, we have considered a spatial decomposition of $f'(\{\phi^{k}\}) $ such that 
\begin{equation}
f'=\sum_{m=1}^{M_f}q^{(m)}_x q^{(m)}_y q^{(m)}_z=\sum_{m=1}^{M_f} \boldsymbol{N}_x \boldsymbol{q}^{(m)}_x \ \boldsymbol{N}_y \boldsymbol{q}^{(m)}_y\ \boldsymbol{N}_z \boldsymbol{q}^{(m)}_z.
\end{equation}
This can be done using the higher-order proper generalized decomposition  \cite{lu2018adaptive,lu2018multi,lu2019datadriven} or higher-order singular value decomposition \cite{bergqvist2010higher} with the data available for $f'(\{\phi^{k}\})$. Each of the three systems of equations corresponds to a one-dimensional problem, whose DoFs can be much smaller than that of the original model \eqref{eq:3.2_stabilized_semi_implicit_discrete}. Orders-of-magnitude reduction in DoFs can be easily achieved for large-scale systems.

\subsection{Incremental subspace mode-updating algorithm}
\label{sec:solution_algorithm}
The three systems of equations in \eqref{eq:3.3_directional_systems} are mutually coupled and can be solved using an alternating iterative scheme. Specifically, we can update each unknown sequentially while holding the other two fixed, and the procedure can be repeated until convergence. To construct the complete TD approximation, we propose the following incremental subspace mode updating algorithm. The overall solution procedure is summarized as follows:

\begin{itemize}
    \item \textbf{Initialization:}
    Compute the first mode of $\boldsymbol{u}$, $\boldsymbol{v}$, and $\boldsymbol{w}$ using the alternating mode solver and initialize the set of retained modes: $\mathcal{B} = \left\{\left(\boldsymbol{u},\boldsymbol{v},\boldsymbol{w}\right)\right\}$.

    \item \textbf{Alternating mode solution:}
     For each new mode, sequentially solve the one-dimensional
    systems \eqref{eq:3.3_directional_systems} for $\boldsymbol{u}$, $\boldsymbol{v}$, and
    $\boldsymbol{w}$ while fixing the other two. Repeat the iterations until convergence or the maximum number of iterations is reached.

   \item \textbf{Subspace mode updating for mode enrichment:} Progressively expand the set of retained modes: $\mathcal{B} \leftarrow \mathcal{B} \cup \{(\boldsymbol{u},\boldsymbol{v},\boldsymbol{w})\}$, until reaching the prescribed subspace basis size $r_{\text{max}}$. Once this size is reached, successively replace the oldest retained modes with newly computed modes. Each updating stage performs up to $N_{\text{rep}}$  mode replacements.

    \item \textbf{Convergence assessment:}
    Evaluate the normalized approximation error after each
    newly computed mode. Repeat the subspace mode updating with an expanded subspace: $r_{\text{max}}\leftarrow r_{\text{max}}+r_{a}$,
    until the prescribed tolerance $\varepsilon$ is satisfied.
\end{itemize}
The complete procedure is summarized in Algorithm~\ref{alg:subspace_update}. Compared with the conventional incremental alternating algorithm employed in proper generalized decomposition  \cite{Chinesta2011}, the proposed algorithm is expected to reduce the number of retained modes through adaptive subspace mode-updating while maintaining computational efficiency. The subspace expansion coefficient $r_a$ and updating coefficient $\alpha_{\mathrm{up}}$ provide flexibility in balancing mode reduction and computational cost. In particular, the conventional incremental alternating algorithm can be recovered as a special case when $r_a=1$ and $\alpha_{\mathrm{up}}=0$. An investigation of different updating coefficients will be conducted in numerical examples.

\begin{algorithm}[htbp]
\caption{Subspace mode updating -- Global loop}
\label{alg:subspace_update}
\begin{algorithmic}[1]
\Require Coefficient matrices, tolerance $\varepsilon$,
updating coefficients $r_a$, $\alpha_{\mathrm{up}}$
\Ensure Retained modes $\mathcal{B}$, number of modes $M$

\State $\mathcal{B} \gets [\textsc{ModeSolver}]$
\State $r_{\max} \gets r_a$, $q \gets 0$,
$N_{\mathrm{rep}} \gets \lceil r_a\alpha_{\mathrm{up}}\rceil$
\State Evaluate convergence error $\eta$

\While{$\eta \geq \varepsilon$}

    \If{$|\mathcal{B}| = r_{\max}$ \textbf{and}
    $q < N_{\mathrm{rep}}$}
        \State Remove the oldest mode from $\mathcal{B}$
        \State $q \gets q+1$
    \ElsIf{$|\mathcal{B}| = r_{\max}$}
        \State $r_{\max} \gets r_{\max}+r_a$, $q \gets 0$
    \EndIf

    \State $M \gets |\mathcal{B}|+1$
    \State Compute a new mode using \textsc{ModeSolver} (Algorithm~\ref{alg:modesolver})
    \State Append the new mode to $\mathcal{B}$
    \State Evaluate the convergence error, e.g.,
    $\eta=\frac{|| \boldsymbol{u}||\ || \boldsymbol{v}||\ || \boldsymbol{w}|| }{\eta_{\mathrm{ref}}} $

\EndWhile

\State \Return $\mathcal{B}$, $M$

\end{algorithmic}
\end{algorithm}

\begin{algorithm}[htbp]
\caption{ModeSolver -- Local loop}
\label{alg:modesolver}
\begin{algorithmic}[1]
\Require 
Coefficient matrices, local tolerance
$\varepsilon_{\mathrm{loc}}$,
maximum iterations $N_{\mathrm{iter}}$
\Ensure Updated mode 
$\boldsymbol{u}$, $\boldsymbol{v}$, $\boldsymbol{w}$

\For{$k=1,\ldots,N_{\mathrm{iter}}$}

    \State Fix $\boldsymbol{v}$ and $\boldsymbol{w}$;
    solve $\boldsymbol{K}_u\ \boldsymbol{u}=\boldsymbol{Q}_u$

    \State Fix $\boldsymbol{u}$ and $\boldsymbol{w}$;
    solve $\boldsymbol{K}_v\ \boldsymbol{v}=\boldsymbol{Q}_v$

    \State Fix $\boldsymbol{u}$ and $\boldsymbol{v}$;
    solve $\boldsymbol{K}_w\ \boldsymbol{w}=\boldsymbol{Q}_w$

    \State Evaluate the local convergence error, e.g.,
    $\eta_{\mathrm{loc}}=\frac{||\Delta \boldsymbol{u}||\ ||\Delta \boldsymbol{v}||\ ||\Delta \boldsymbol{w}|| }{\eta_{\mathrm{loc, ref}}}$

    \If{$\eta_{\mathrm{loc}} < \varepsilon_{\mathrm{loc}}$}
        \State \textbf{break}
    \EndIf

\EndFor

\State \Return
$\boldsymbol{u}$, $\boldsymbol{v}$, $\boldsymbol{w}$

\end{algorithmic}
\end{algorithm}

\subsection{GPU implementation}
To further accelerate large-scale grain growth
simulations, the proposed TD framework is implemented
on a GPU workstation to exploit parallelism across multiple  OPs. We present here two key strategies that are
developed for efficient GPU implementation: GPU-resident flattened solution storage
and a batched TD solver. These strategies aim to
improve computational efficiency by facilitating
parallel execution, reducing memory-management
overhead, and minimizing host-device data transfers.
\\
\\
\noindent\textbf{GPU-resident flattened solution storage.}
The object-based data structure used in the CPU implementation
is less suitable for GPU execution, as managing numerous
independent grain objects can lead to irregular memory access
and frequent memory allocation. To facilitate efficient GPU
execution, the grain data are reorganized into contiguous
GPU-resident arrays. For each spatial direction, the one-dimensional TD modes of all OPs are stored in a flattened array.
For example, the $x$-direction modes of the $i$-th OP are
first concatenated into a one-dimensional vector
\begin{equation}
\label{eq:3.6_flattened_mode_storage}
\boldsymbol{U}_{x,i}
= \operatorname{col}\left(
\boldsymbol{u}_i^{(1)},\ldots,
\boldsymbol{u}_i^{(M_i)}
\right),
\end{equation}
where $M_i$ denotes the number of retained modes for the $i$-th OP. These mode vectors are further concatenated into a global array for all the OPs
\begin{equation}
\boldsymbol{U}_x
= \operatorname{col}\left(
\boldsymbol{U}_{x,1},\ldots,
\boldsymbol{U}_{x,N}
\right).
\end{equation}
An offset array records the starting position of each
grain block, allowing efficient access to grain-specific
data despite variations in subdomain sizes and mode counts.
The same storage strategy is adopted for the $y$- and
$z$-direction modes. Additional information, including
bounding boxes of OPs, local subdomain dimensions, mode counts,
activity flags, and neighbor lists, is also maintained
on the GPU. 

This flattened representation facilitates contiguous
memory storage and enables multiple grains to be processed
collectively using GPU kernels. Furthermore, keeping
the solution data resident on the GPU throughout time
integration reduces repeated host-device data transfers
between consecutive time steps.
\\
\\
\textbf{Batched TD solver}. As the TD method reduces the 3D problem  to a sequence of small 1D systems, the systems of equations for different OPs  can be organized into contiguous arrays and processed collectively using batched GPU kernels. Grain-specific offsets and local dimensions are used to identify individual systems within these arrays. For each spatial direction,  only the sparse coefficients of matrices are generated directly in the flattened GPU arrays. The resulting independent systems can be solved concurrently using a batched Thomas algorithm. Although the Thomas algorithm is inherently sequential within each system, parallel execution across multiple OPs enables efficient GPU utilization. The batched solver is applied sequentially to the $\boldsymbol{u}$-, $\boldsymbol{v}$-, and $\boldsymbol{w}$-direction updates, preserving the alternating solution structure while enabling concurrent processing of multiple OPs on the GPU.

\section{Numerical Experiments}

\subsection{3D grain growth simulations}
\label{sec:3d_grain_growth}
We use 3D grain growth simulations to evaluate the accuracy and computational performance of the proposed TD framework. A domain size of \(96\,\upmu\mathrm{m}\times76.8\,\upmu\mathrm{m}\times38.4\,\upmu\mathrm{m}\) is considered first. The initial grain structure is generated using centroidal Voronoi tessellation (CVT) followed by Lloyd's algorithm and an initialization using a signed-distance function for the diffusive representation of grain boundaries. Each grain is represented by a distinct OP, with a unique identifier and an assigned crystallographic orientation.  The physical and kinetic parameters used in the simulations are summarized in Table~\ref{tab:4.1_physical_parameters} and are representative  for 316L stainless steel.

First,  a mesh size of $300\times240\times120$ elements is  applied to the domain with 100 grains with randomly assigned grain orientations and OPs. The simulation is conducted  with $\Delta t=5\times10^{-5}\,\mathrm{s}$ for 500 time steps, corresponding to a total simulation time of $25\,\mathrm{ms}$. We use both FEM and TD under the same settings, with the stabilized semi-implicit time integration scheme, for the phase-field simulation. The hardware used for this work is the Intel Xeon Gold 6548Y+ Processor with 32 CPU cores and 512 GB RAM and a NVIDIA Quadro RTX A6000 GPU. 

\begin{table}[H]
\centering
\caption{Physical parameters used in the grain growth simulations \cite{yuan2026efficient}.}
\begin{tabular}{ll}
\toprule
\textbf{Parameter} & \textbf{Value} \\
\midrule
$\zeta$ & $2.5\Delta x$ \\
$H_{\mathrm{latent}}$ & $1.9\times10^{-9}\ \mathrm{J}/\upmu\mathrm{m}^{3}$ \\
$\gamma_{SL}$ & $2\times10^{-13}\ \mathrm{J}/\upmu\mathrm{m}^{2}$ \\
$W$ & $6\gamma_{SL}/\zeta$ \\
$\kappa$ & $9\gamma_{SL}\zeta/2$ \\
$\mu_0$ & $2.17\times10^{5}\ \upmu\mathrm{m}/(\mathrm{s}\cdot\mathrm{K})$ \\
$T_{\mathrm{Liq}}$ & $1700\ \mathrm{K}$ \\
$L$ & $\mu_0T_{\mathrm{Liq}}/(150\zeta H_{\mathrm{latent}})$ \\
\bottomrule
\end{tabular}
\label{tab:4.1_physical_parameters}
\end{table}

\begin{figure}[htbp]
  \centering
  \subcaptionbox{FEM, \,$t=0\,\mathrm{ms}$}[0.25\textwidth]{%
    \includegraphics[width=\linewidth]{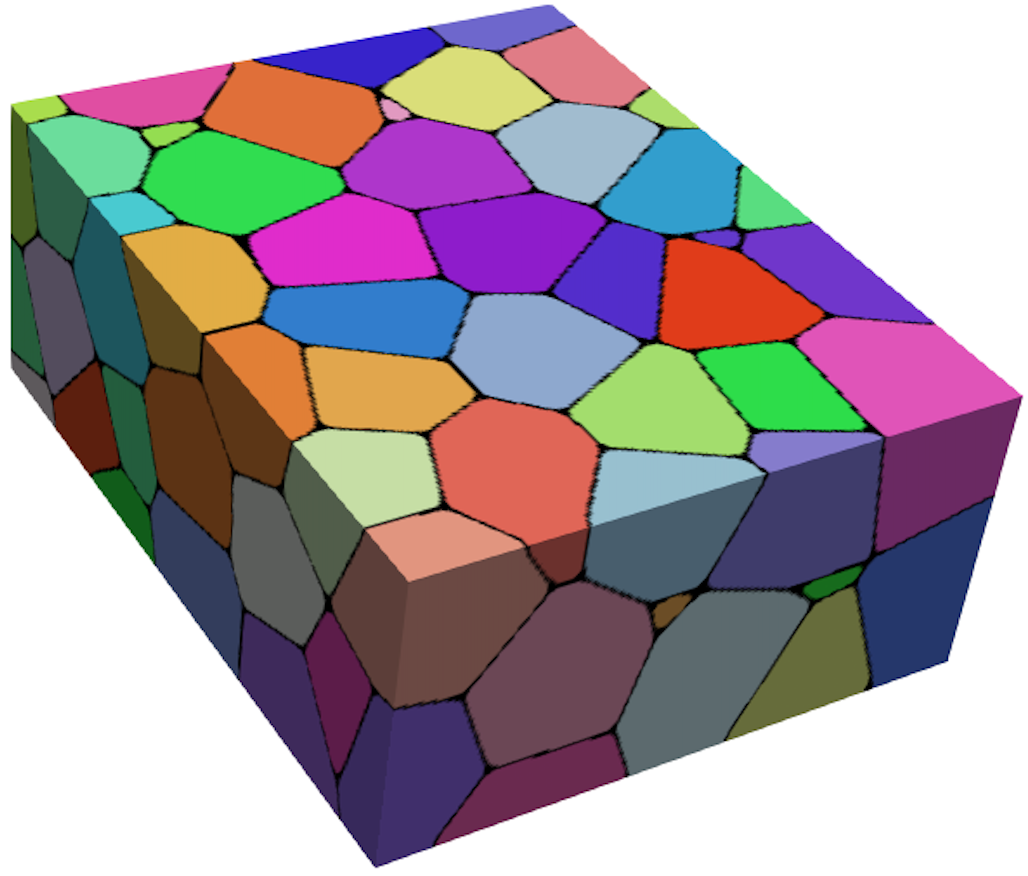}
  }
  \hspace{0.05\textwidth}
  \subcaptionbox{FEM, \,$t=12.5\,\mathrm{ms}$}[0.25\textwidth]{%
    \includegraphics[width=\linewidth]{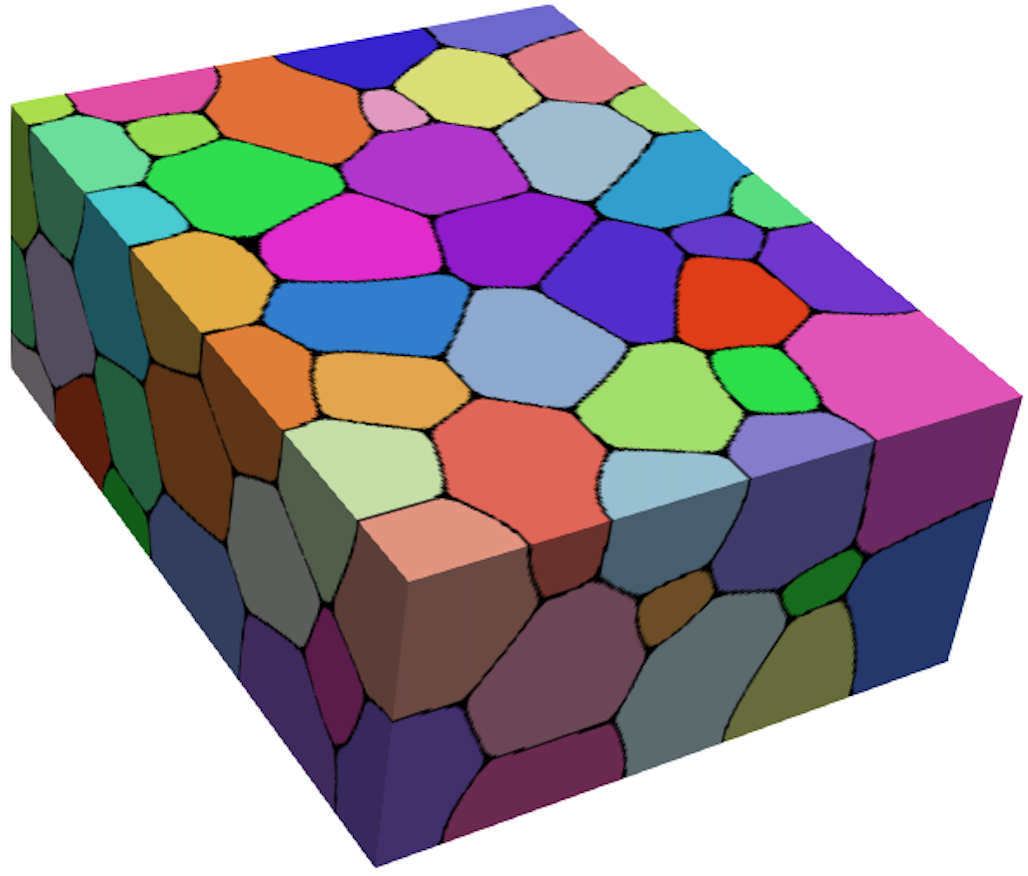}
  }
  \hspace{0.05\textwidth}
  \subcaptionbox{FEM, \,$t=25\,\mathrm{ms}$}[0.25\textwidth]{%
    \includegraphics[width=\linewidth]{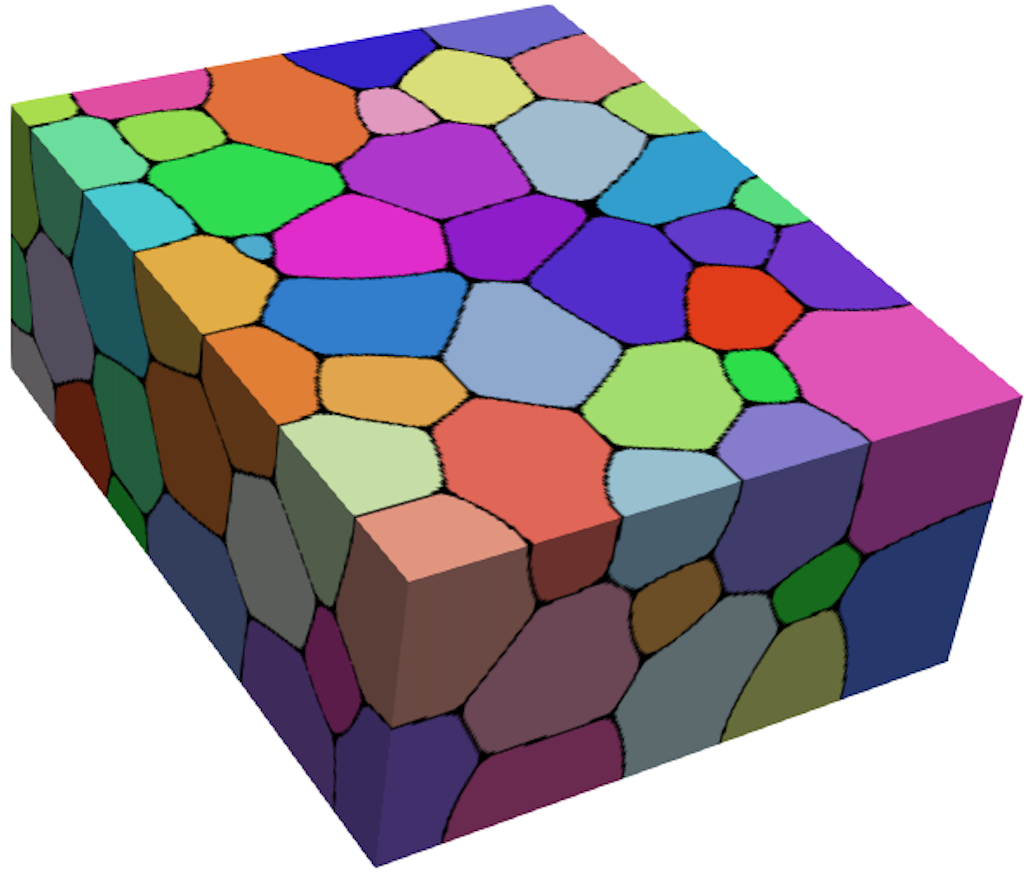}
  }

  \vspace{0.02\textwidth}

  \subcaptionbox{TD, \,$t=0\,\mathrm{ms}$}[0.25\textwidth]{%
    \includegraphics[width=\linewidth]{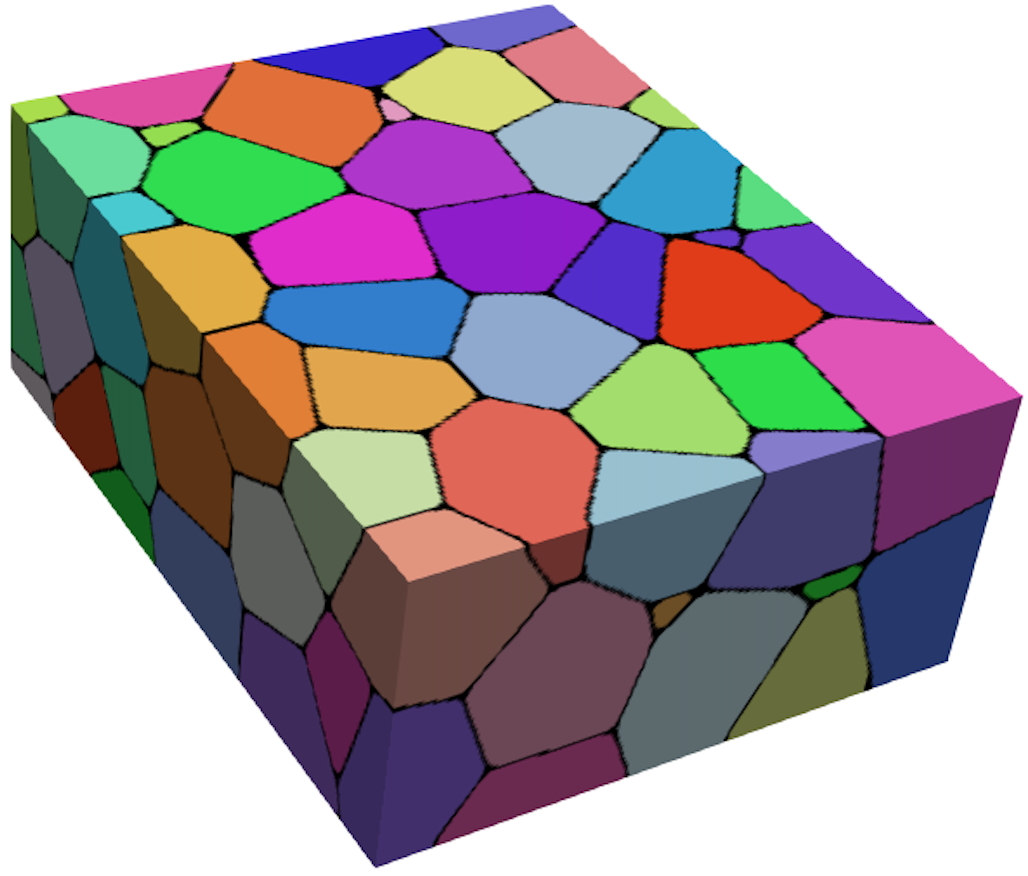}
  }
  \hspace{0.05\textwidth}
  \subcaptionbox{TD, \,$t=12.5\,\mathrm{ms}$}[0.25\textwidth]{%
    \includegraphics[width=\linewidth]{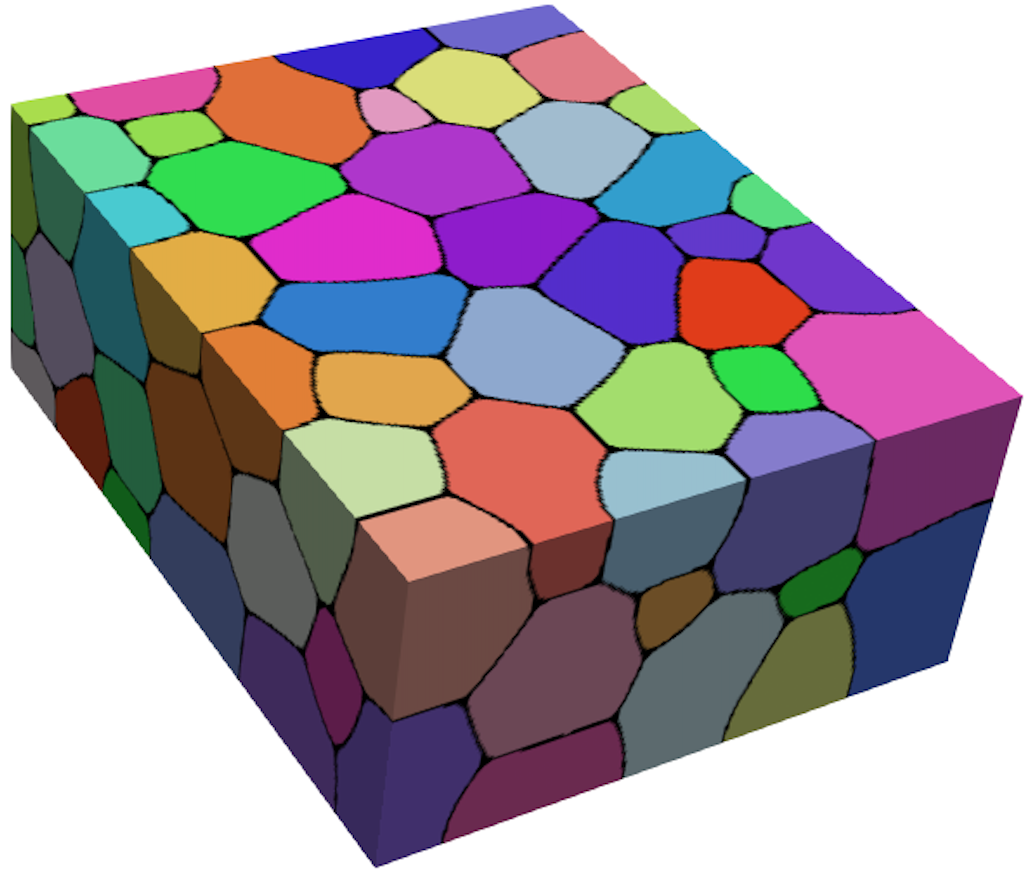}
  }
  \hspace{0.05\textwidth}
  \subcaptionbox{TD, \,$t=25\,\mathrm{ms}$}[0.25\textwidth]{%
    \includegraphics[width=\linewidth]{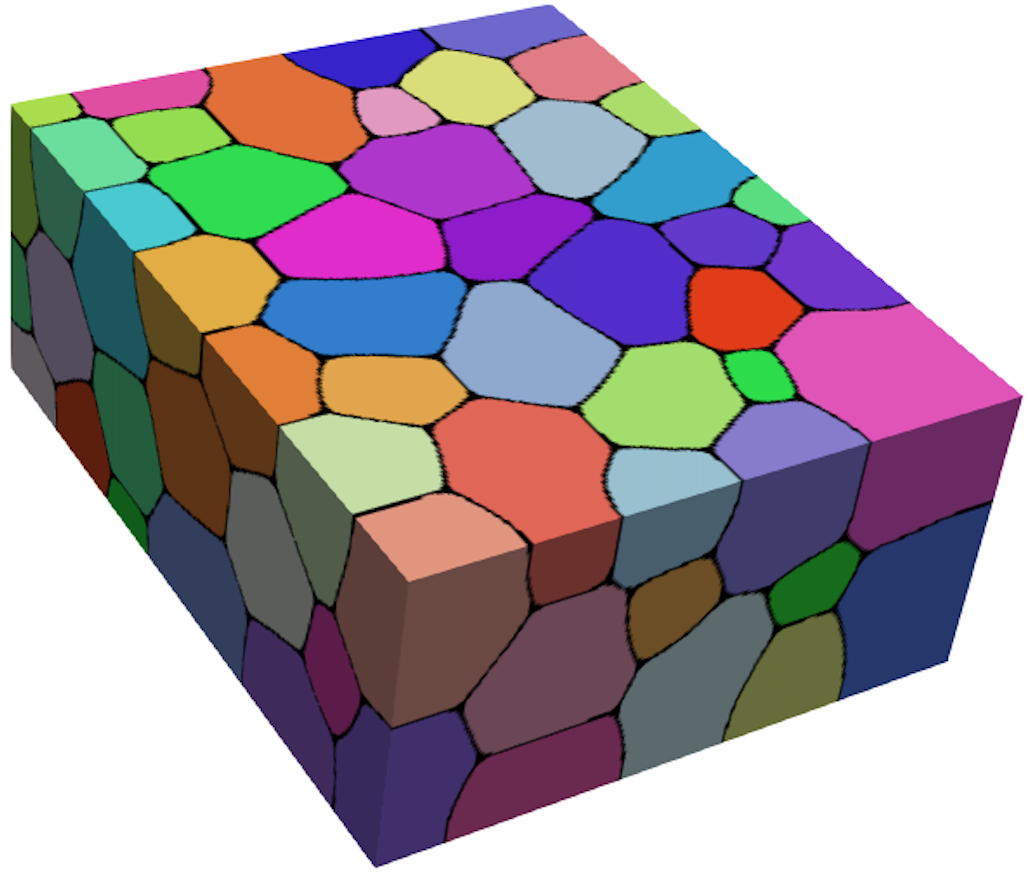}
  }

  \caption{Comparison of the 3D grain growth obtained using FEM and TD: (a)-(c) FEM and (d)-(f) TD.}
  \label{fig:4.1_cpu_grain_evolution}
\end{figure}

Figure~\ref{fig:4.1_cpu_grain_evolution} compares the grain structures predicted by the FEM and TD methods at three time steps, $t=0$, $12.5$, and $25\, \mathrm{ms}$. Both methods produce highly similar grain structures and exhibit consistent grain coarsening behaviors. As time progresses, grain boundaries migrate, leading to changes in grain size and morphology as the microstructure evolves toward a lower-energy configuration. In particular, smaller grains (e.g., the green one) tend to shrink or disappear, while neighboring grains (e.g., the pink one at the corner) grow through grain boundary migration. The TD solution closely reproduces these microstructure changes observed in the reference FEM results, including the evolution of individual grain morphologies and the overall grain boundary network. The qualitative agreement between the two methods throughout the simulation demonstrates the capability of the TD formulation to capture the complex dynamics of grain evolution while representing the full 3D solution through low-rank structures.

\begin{figure}[htbp]
  \centering
  \subcaptionbox{\,$t=0\,\mathrm{ms}$}[0.28\textwidth]{%
    \includegraphics[width=\linewidth]{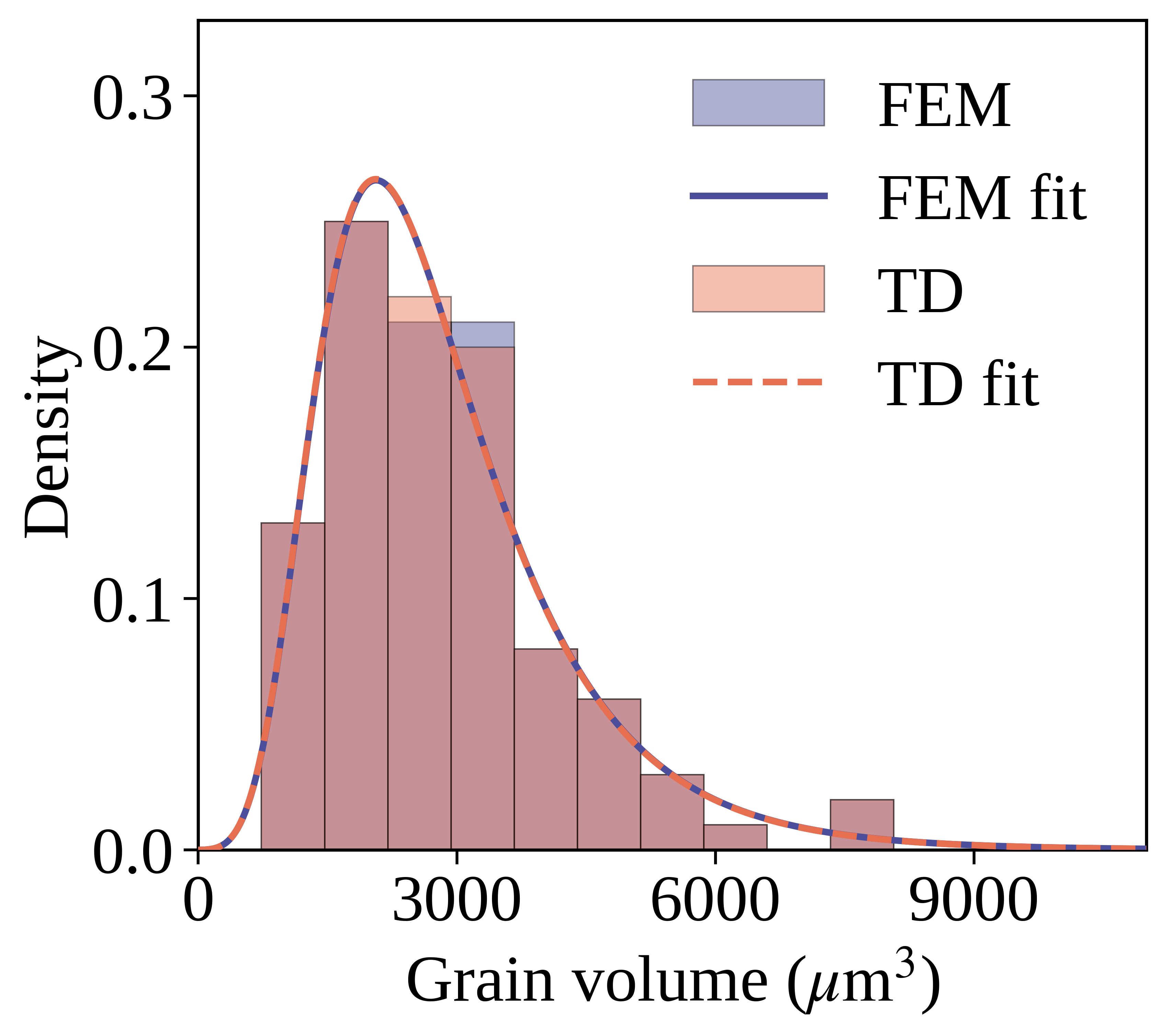}
  }
  \hspace{0.05\textwidth}
  \subcaptionbox{\,$t=12.5\,\mathrm{ms}$}[0.28\textwidth]{%
    \includegraphics[width=\linewidth]{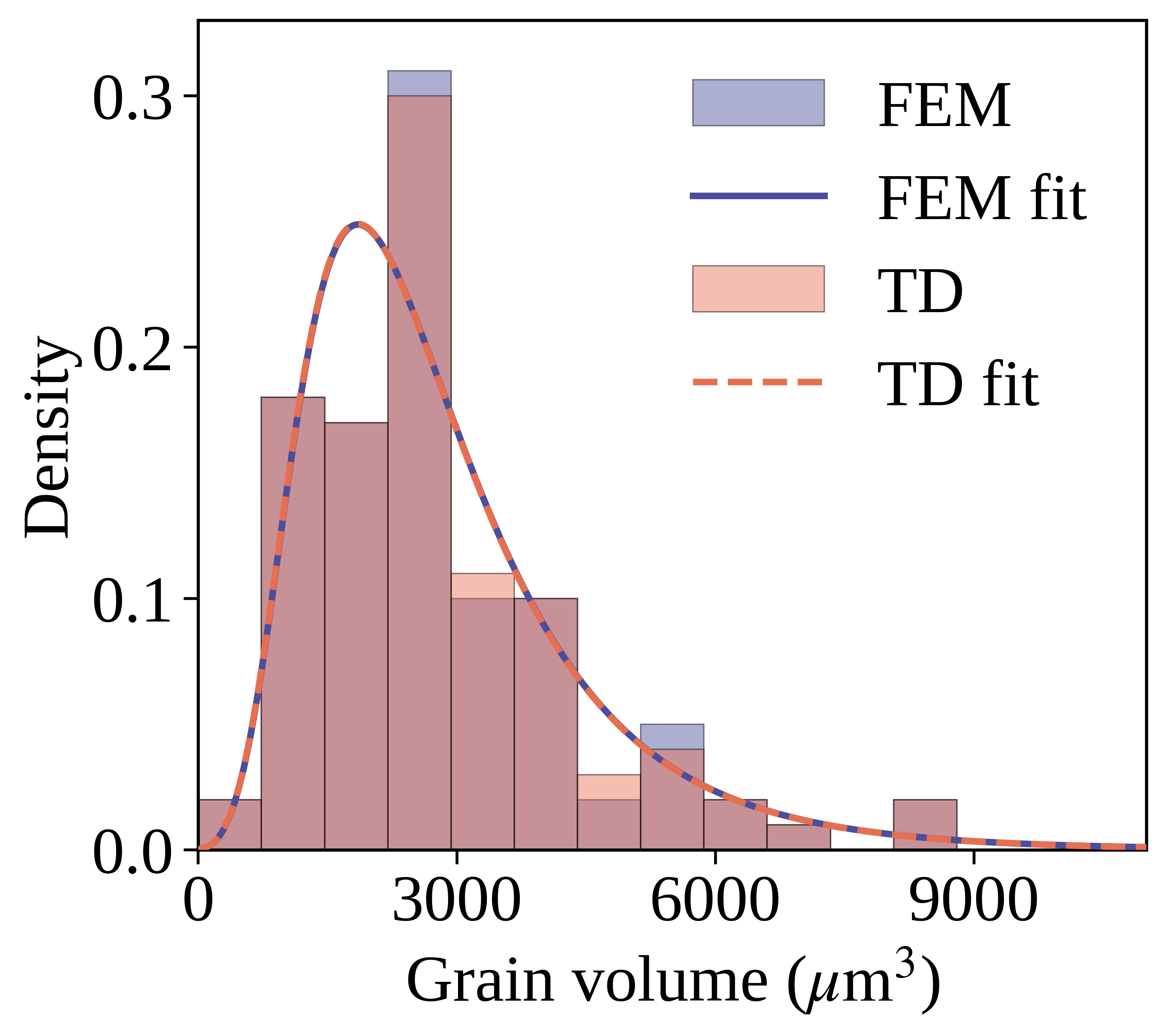}
  }
  \hspace{0.05\textwidth}
  \subcaptionbox{\,$t=25\,\mathrm{ms}$}[0.28\textwidth]{%
    \includegraphics[width=\linewidth]{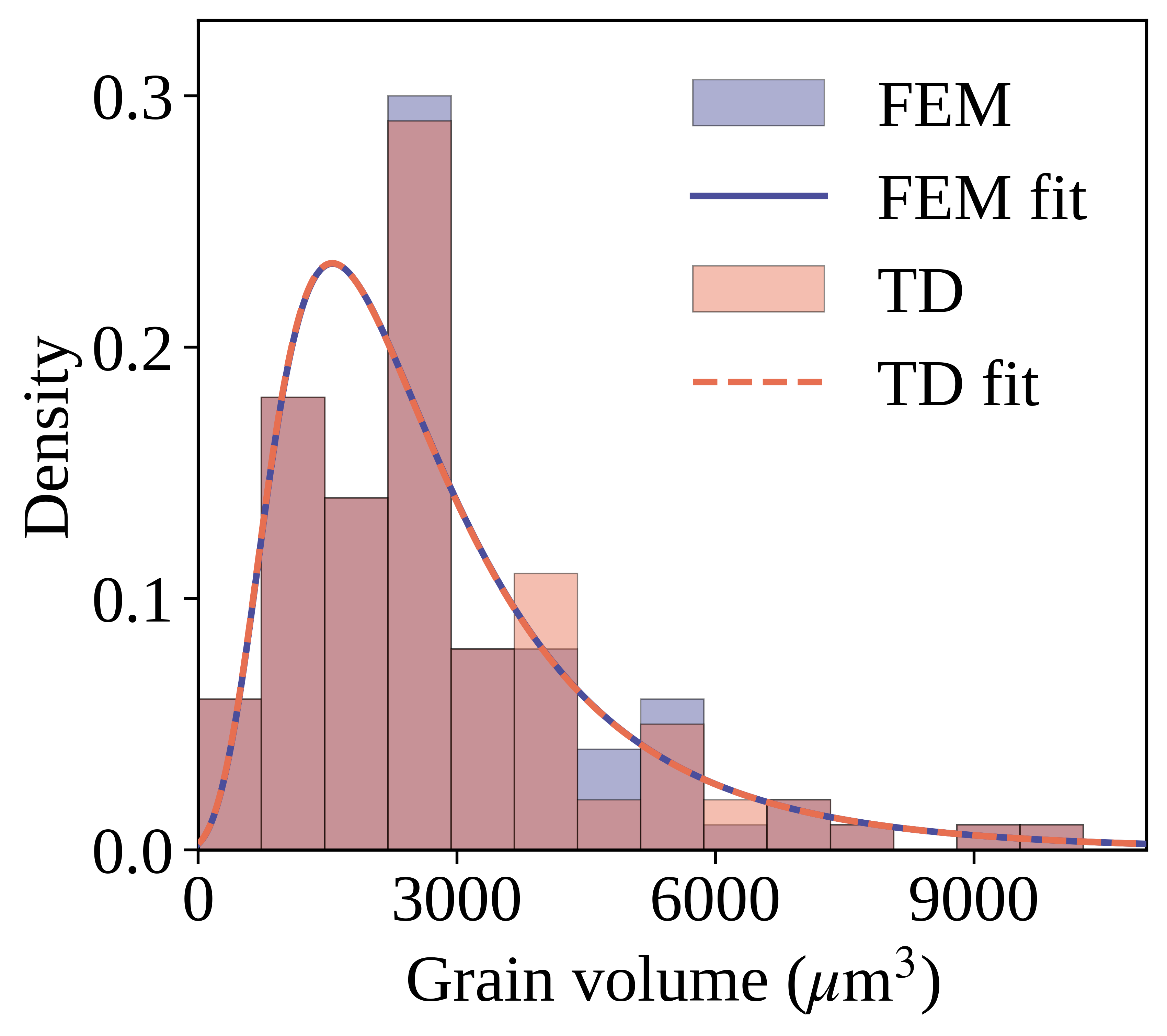}
  }

  \vspace{0.02\textwidth}

  \subcaptionbox{\,$t=0\,\mathrm{ms}$}[0.28\textwidth]{%
    \includegraphics[width=\linewidth]{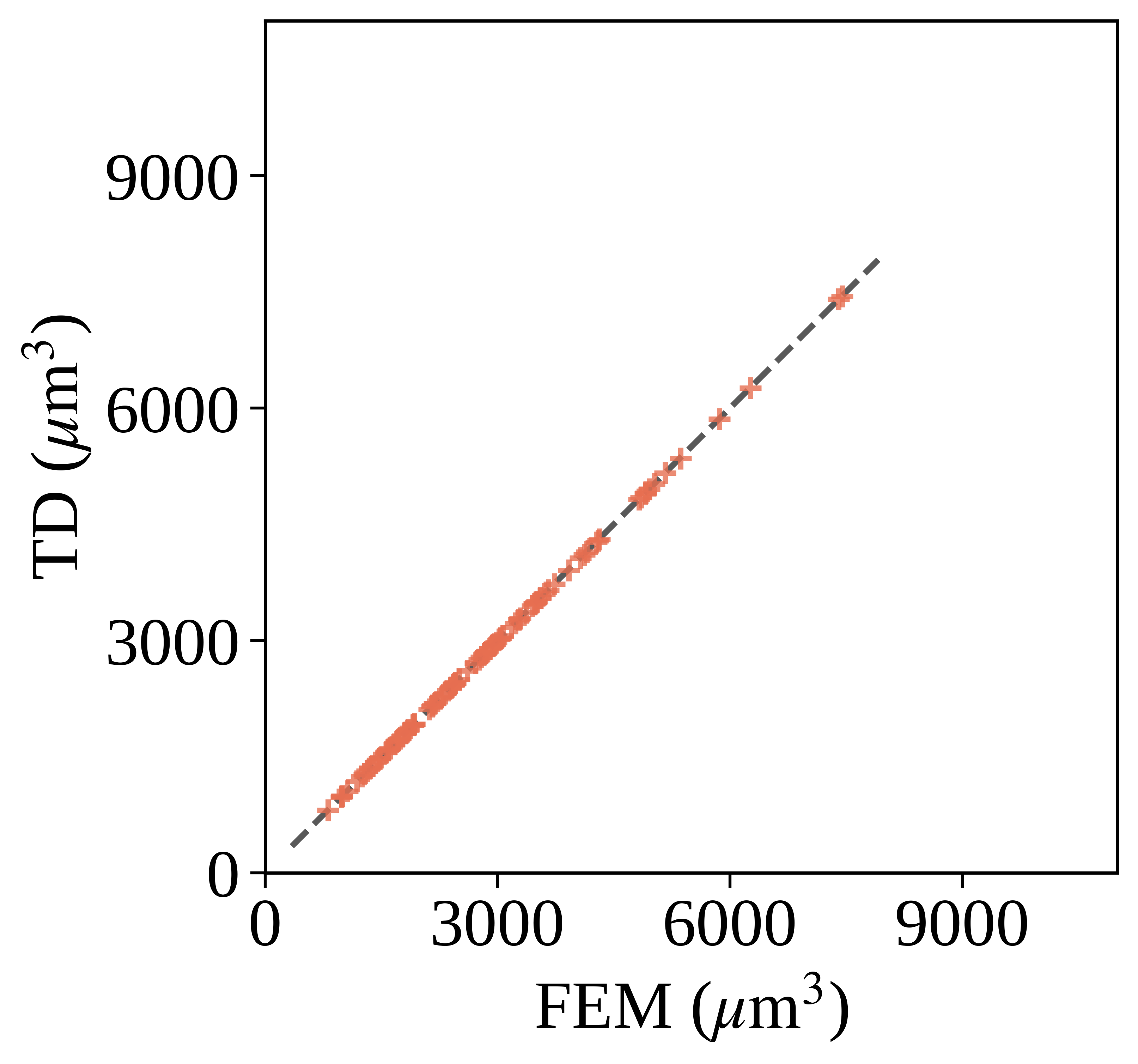}
  }
  \hspace{0.05\textwidth}
  \subcaptionbox{\,$t=12.5\,\mathrm{ms}$}[0.28\textwidth]{%
    \includegraphics[width=\linewidth]{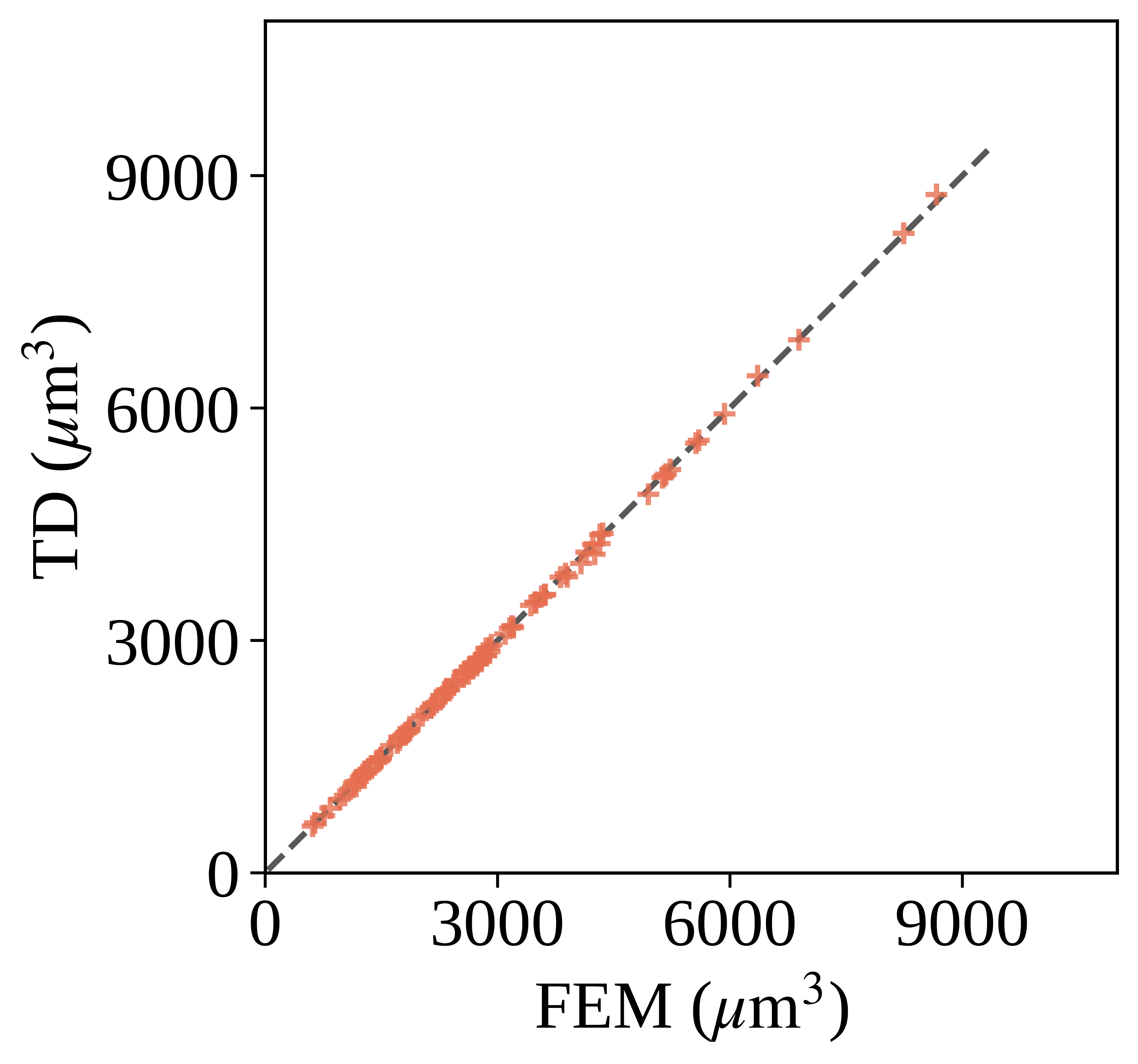}
  }
  \hspace{0.05\textwidth}
  \subcaptionbox{\,$t=25\,\mathrm{ms}$}[0.28\textwidth]{%
    \includegraphics[width=\linewidth]{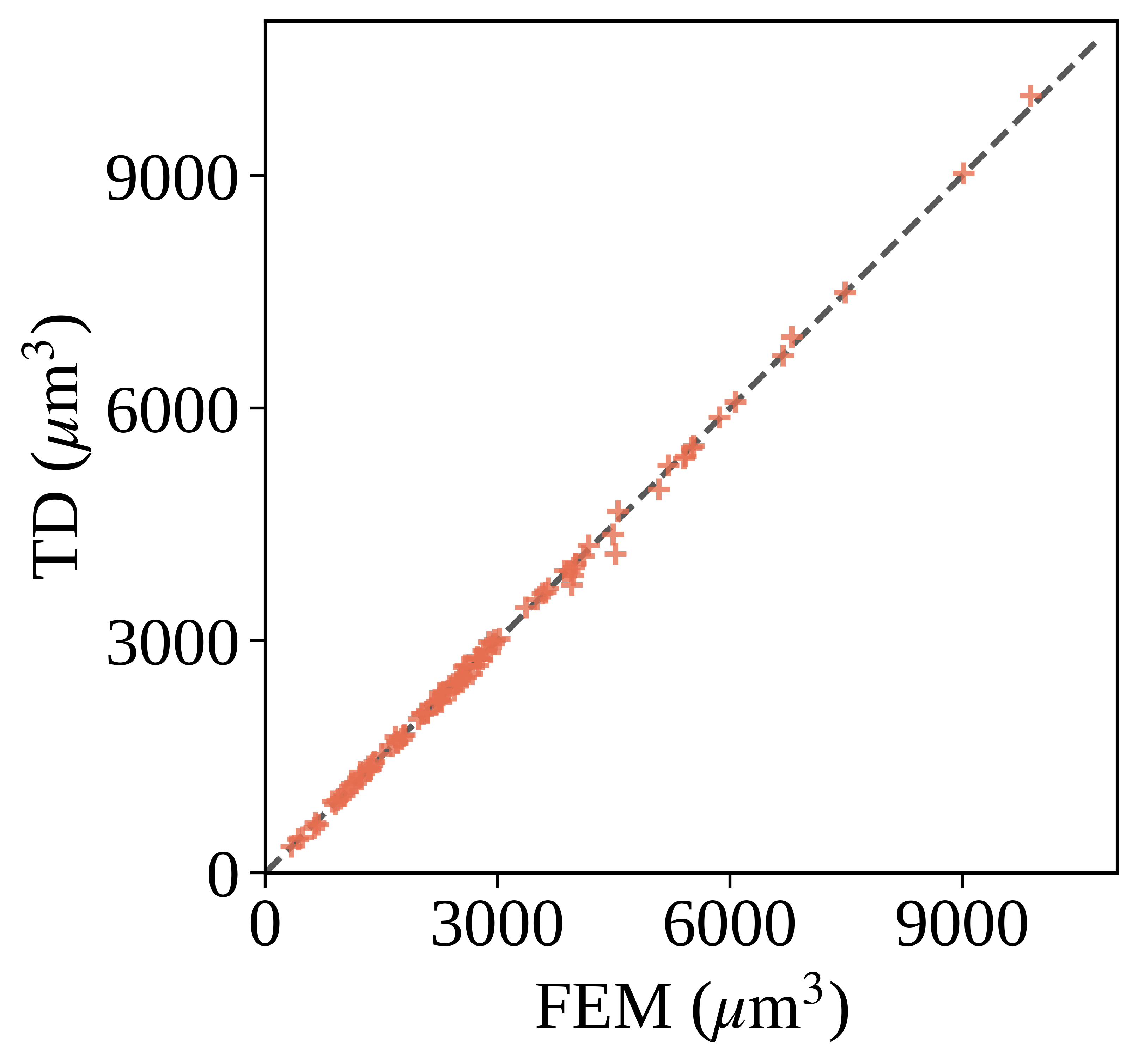}
  }

  \caption{Grain volume comparison between FEM and TD: (a)-(c) volume distributions with lognormal fits and (d)-(f) grain-to-grain volume comparison, with MARE $=$ 0.16\%, 0.84\%, and 1.80\%. $\mathrm{MARE}=\frac{1}{N}\sum_{i=1}^{N}\left|\frac{V_i^{\mathrm{TD}}-V_i^{\mathrm{FEM}}}{V_i^{\mathrm{FEM}}}\right|\times100\%$.}
  \label{fig:4.1_cpu_grain_volume}
\end{figure}

Figure~\ref{fig:4.1_cpu_grain_volume} presents a quantitative comparison of individual grain volumes predicted by the FEM and TD methods. Figures~\ref{fig:4.1_cpu_grain_volume}(a)--(c) show that both methods yield nearly identical grain volume distributions and corresponding lognormal fits. As the time progresses, the distributions become progressively broader, reflecting the characteristic coarsening behavior in which larger grains grow at the expense of smaller ones. Figures~\ref{fig:4.1_cpu_grain_volume}(d)--(f) further compare the volumes of individual grains predicted by the two methods. The data points are closely clustered around the reference line, indicating excellent agreement between the FEM and TD results. The corresponding mean absolute relative errors (MAREs) are only 0.16\%, 0.84\%, and 1.80\% at $t=0$, $12.5$, and $25,\mathrm{ms}$, respectively. These results demonstrate that the TD method accurately reproduces both the statistical characteristics of grain volume distributions and the evolution of individual grains.

\begin{figure}[htbp]
  \centering
  \subcaptionbox{$r_{a}=1,\;\alpha_{\text{up}}=0$}[0.35\textwidth]{%
    \includegraphics[width=\linewidth]{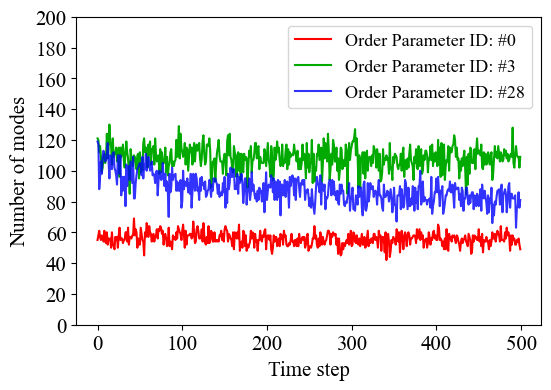}
  }
  \hspace{0.05\textwidth}
  \subcaptionbox{$r_{a}=3,\;\alpha_{\text{up}}=1$}[0.35\textwidth]{%
    \includegraphics[width=\linewidth]{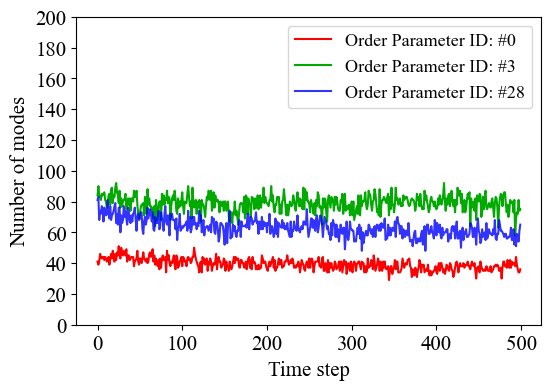}
  }
  \caption{Evolution of the number of modes for three representative OPs.}
  \label{fig:4.1_mode_evolution}
\end{figure}

To confirm the effectiveness of the TD approximation, Figure~\ref{fig:4.1_mode_evolution} shows the evolution of the number of modes over time, using different subspace mode-updating parameters $r_{a}$ and $\alpha_{\text{up}}$. For $r_{a}=1$ and $\alpha_{\text{up}}=0$, corresponding to the conventional incremental alternating algorithm, the number of modes remains relatively stable  at approximately 100 throughout the simulation. With $r_{a}=3$ and $\alpha_{\text{up}}=1$, much fewer modes are needed under the same convergence criterion. These results demonstrate the effectiveness of the proposed subspace mode-updating in reducing the number of modes for a compact TD representation of the evolving grain microstructure.

\begin{figure}[htbp]
  \centering

  \subcaptionbox{OP \#3, $r_{a}=1,\;\alpha_{\text{up}}=0$}[0.35\textwidth]{%
    \includegraphics[width=\linewidth]{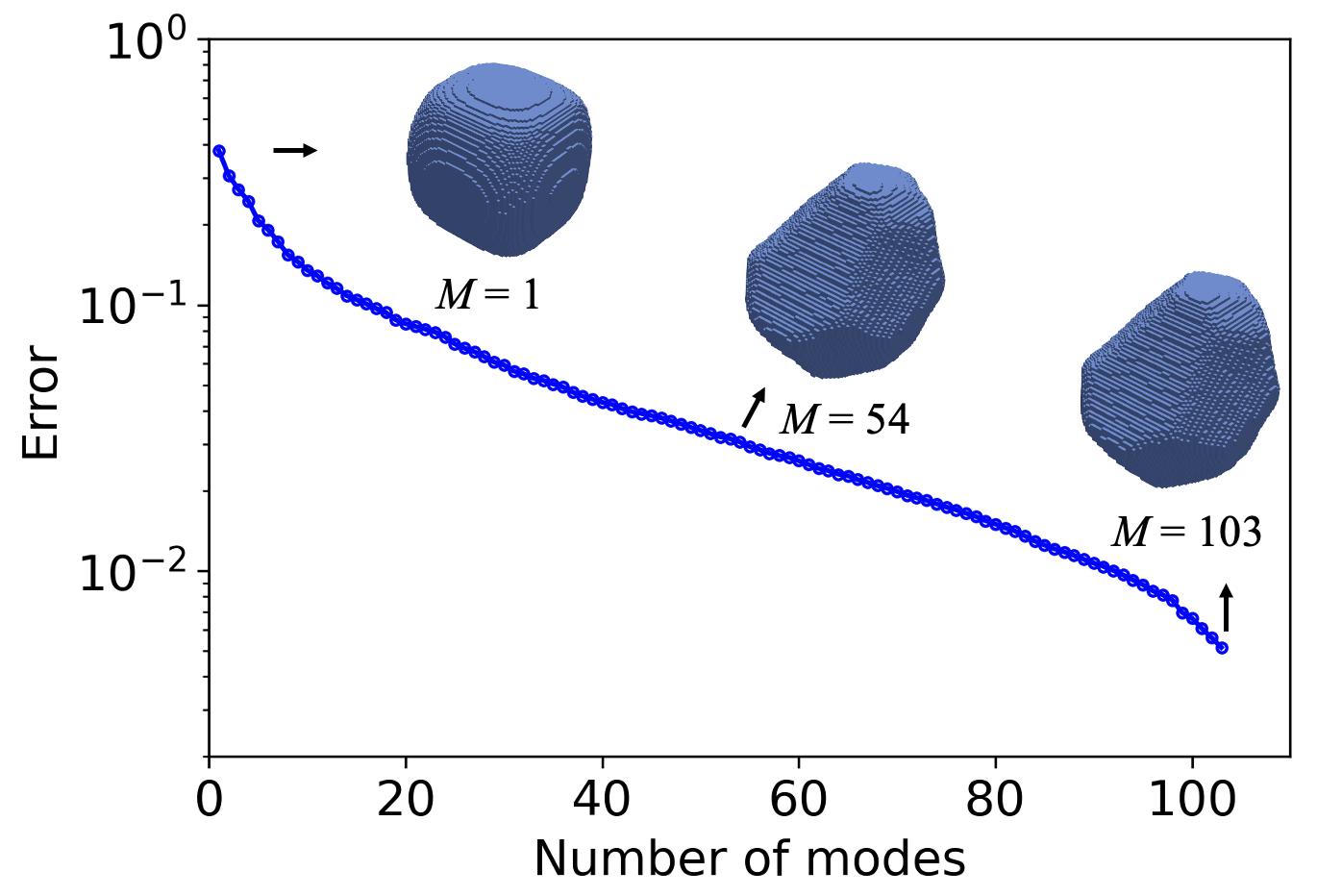}
  }
  \hspace{0.05\textwidth}
  \subcaptionbox{OP \#28, $r_{a}=1,\;\alpha_{\text{up}}=0$}[0.35\textwidth]{%
    \includegraphics[width=\linewidth]{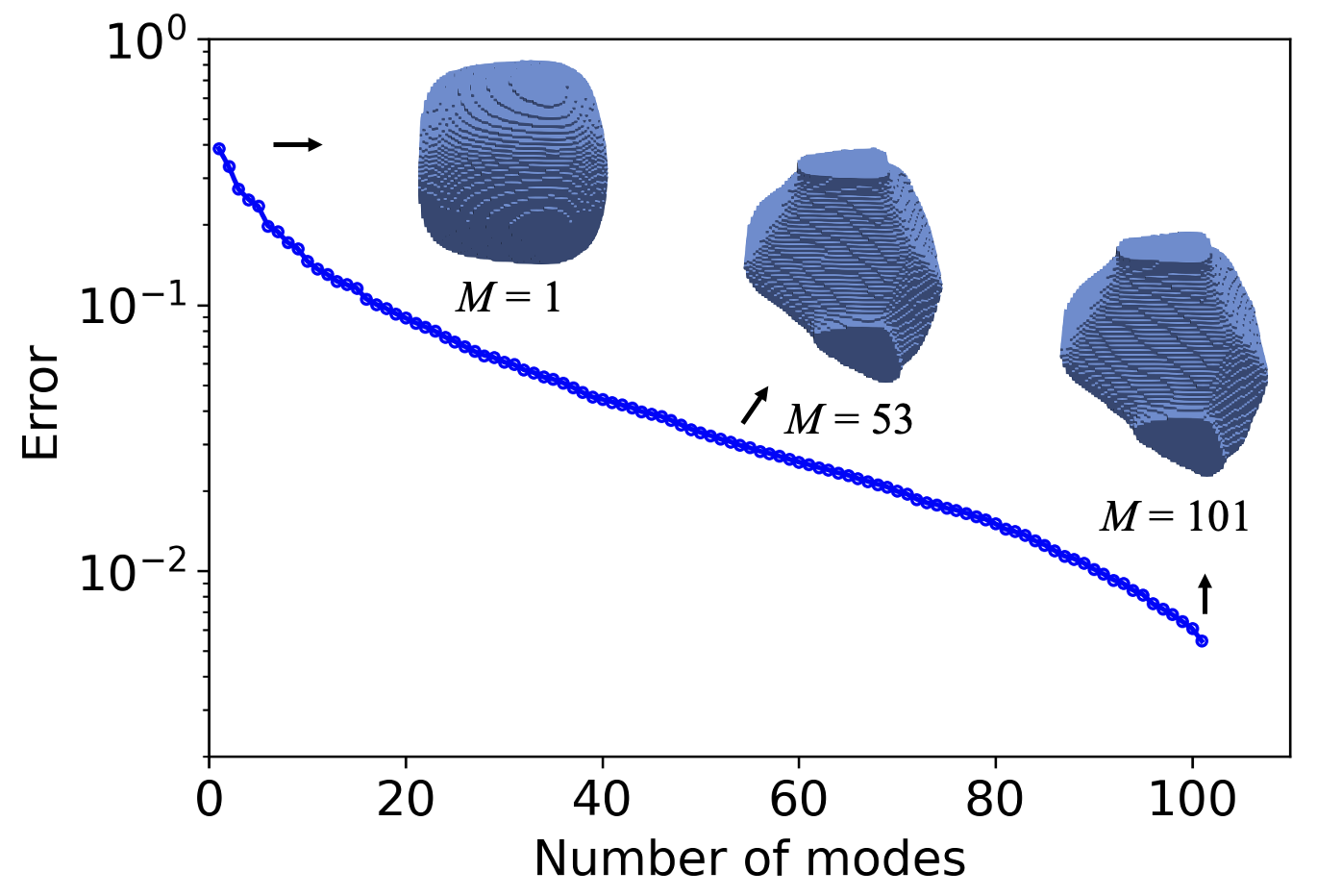}
  }

  \vspace{0.02\textwidth}

  \subcaptionbox{OP \#3, $r_{a}=3,\;\alpha_{\text{up}}=1$}[0.35\textwidth]{%
    \includegraphics[width=\linewidth]{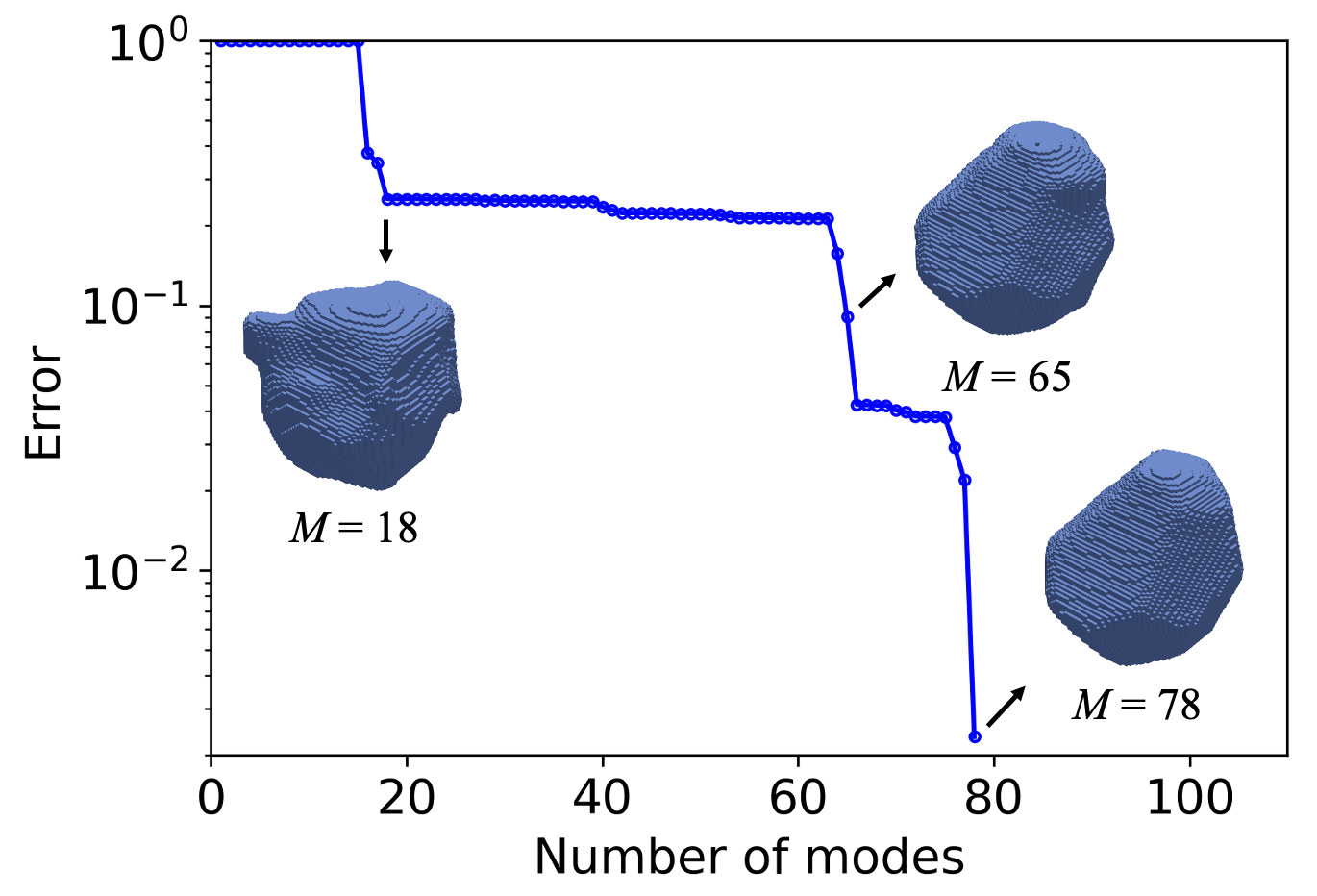}
  }
  \hspace{0.05\textwidth}
  \subcaptionbox{OP \#28, $r_{a}=3,\;\alpha_{\text{up}}=1$}[0.35\textwidth]{%
    \includegraphics[width=\linewidth]{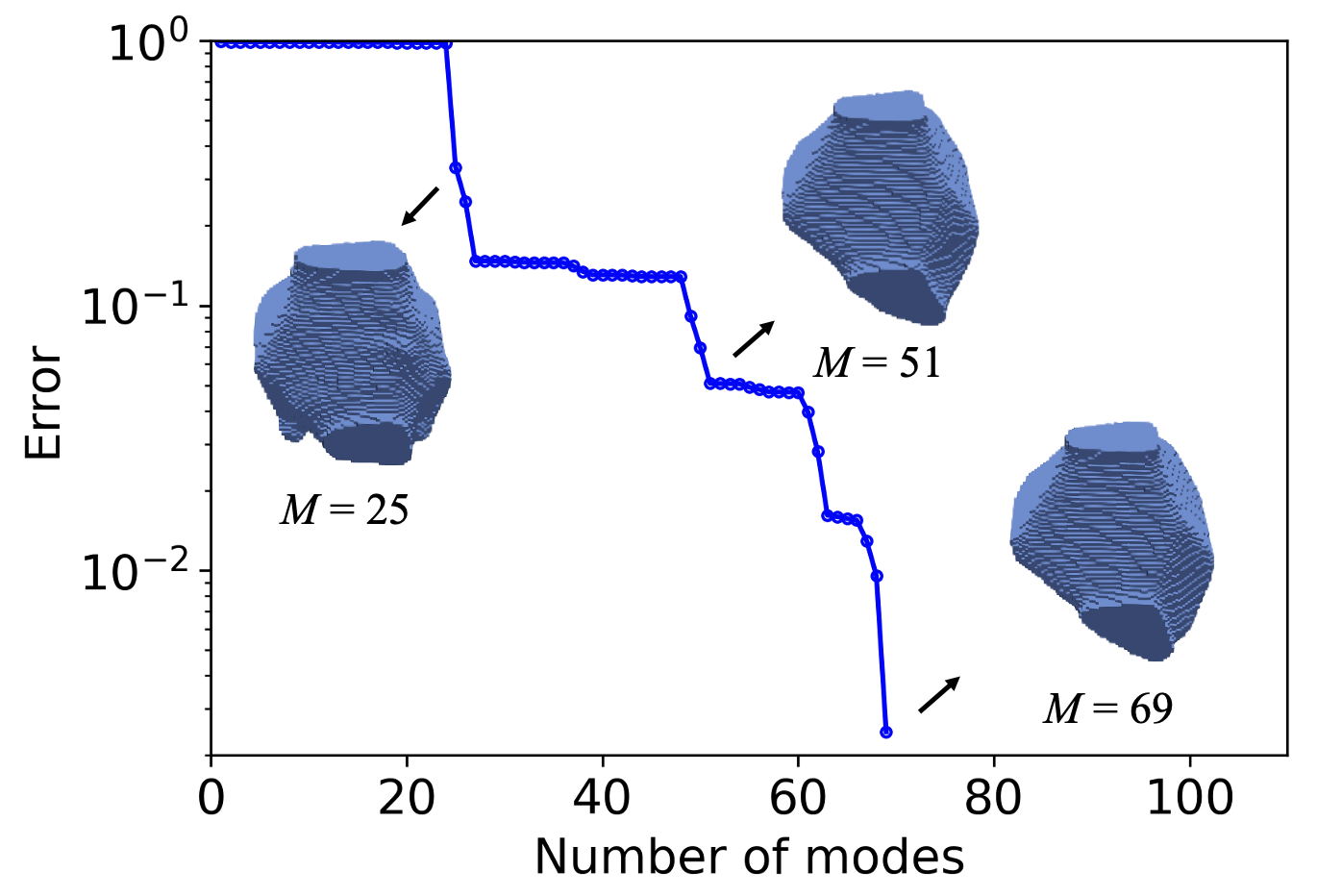}
  }

  \caption{Truncation error of the TD representation with respect to an increasing number of  modes for different subspace mode-updating parameters.}
  \label{fig:4.1_truncation_error}
\end{figure}

To examine the convergence behavior of the TD modes at a given time step, we calculate the truncation error with  different numbers of modes by taking their difference with respect to the final representation. The results for OP \#3 and \#28 are illustrated in Figure~\ref{fig:4.1_truncation_error}. As we can see, both results show favorable convergence behavior.
For \(r_{a}=1\) and \(\alpha_{\text{up}}=0\), the modes are generated successively without replacement. The leading modes therefore contain the dominant contributions, and the truncation error decreases progressively as more modes are included. In contrast, with \(r_{a}=3\) and \(\alpha_{\text{up}}=1\),  the subspace mode-updating algorithm successively replaces previously generated modes, redistributing the solution contributions among the retained modes. As a result, later modes can contribute more significantly to the final approximation, rather than the dominant contributions being concentrated in the leading modes. This behavior highlights the ability of subspace mode updating to refine the TD approximation through mode replacement, facilitating a more compact representation of the evolving grain structure.

A systematic examination of different mode-updating parameters are illustrated in Table~\ref{tab:4.1_subspace_performance}, where we compare resulting average number of modes, data size, run-time overhead. As wee can see, increasing $\alpha_{\text{up}}$ can effectively reduce the number of modes and the data size, but at the expense of increased computational cost. The influence of $r_{a}$ is relatively modest for the cases considered here. Based on these observations, $r_a=3$ and $\alpha_{\mathrm{up}}=1$ are therefore selected for the subsequent simulations. Compared with the conventional incremental alternating algorithm ($r_{a}=1$ and $\alpha_{\text{up}}=0$), this configuration reduces the average number of retained modes from 85 to 62 and the data size from 0.2695~TB to 0.1920~TB, corresponding to reductions of approximately 27\% and 29\%, respectively, with only 19\% increase in runtime,  providing a reasonable balance between compression and computational cost. This demonstrates that the subspace mode-updating algorithm can achieve a favorable trade-off between representation compactness and computational efficiency.

\begin{table}[h]
\centering
\caption{Performance of the TD method with different subspace mode-updating parameters.}
\small
\begin{tabular}{cccc}
\toprule
\textbf{Method} & \textbf{Average Modes} & \textbf{Data Size (TB)} & \textbf{Run-Time Overhead} \\
\midrule
TD ($r_{a}=1$, $\alpha_{\text{up}}=0$) & 85 & 0.2695 & -- \\
TD ($r_{a}=3$, $\alpha_{\text{up}}=1$) & 62 & 0.1920 & 19\% \\
TD ($r_{a}=3$, $\alpha_{\text{up}}=2$) & 56 & 0.1707 & 34\% \\
TD ($r_{a}=3$, $\alpha_{\text{up}}=3$) & 53 & 0.1622 & 47\% \\
TD ($r_{a}=3$, $\alpha_{\text{up}}=5$) & 49 & 0.1526 & 82\% \\
TD ($r_{a}=5$, $\alpha_{\text{up}}=1$) & 62 & 0.1924 & 17\% \\
TD ($r_{a}=5$, $\alpha_{\text{up}}=2$) & 56 & 0.1747 & 33\% \\
TD ($r_{a}=5$, $\alpha_{\text{up}}=3$) & 53 & 0.1635 & 52\% \\
TD ($r_{a}=5$, $\alpha_{\text{up}}=5$) & 50 & 0.1542 & 82\% \\
\bottomrule
\end{tabular}
\label{tab:4.1_subspace_performance}
\end{table}

Finally, we evaluate the computational performance of the FEM and TD methods with increasing mesh resolution. To assess their performance over extended simulations, all cases are run for 10,000 time steps with $\Delta t=5\times10^{-6}\ \mathrm{s}$. For both methods, all the matrix coefficients are computed in parallel on the GPU, while the linear systems are solved on CPUs using vectorized operations. Table~\ref{tab:4.1_cpu_performance} summarizes the total runtime and solution data size  of the two methods. The TD method consistently outperforms FEM across all three cases, with the speedup increasing from 3.08$\times$ at 8.78 million DoFs to 8.35$\times$ at 69.7 million DoFs. Meanwhile, the storage required by the TD representation is as low as 0.61\% of that required by FEM. These results demonstrate that the TD method offers substantial advantages in computational efficiency and data size, which become increasingly pronounced as the problem size grows.

\begin{table}[h]
\centering
\caption{Performance comparison of FEM and TD on CPUs.}
\small
\begin{tabular}{ccccccc}
\toprule
\textbf{Mesh Size} &
\textbf{Total DoFs} &
\textbf{Method} &
\textbf{Total Time (h)} &
\textbf{Speedup} &
\textbf{Data Size (TB)} &
\textbf{Relative Data Size} \\
\midrule
\multirow{2}{*}{$300\times240\times120$}
& \multirow{2}{*}{8.78M}
& FEM & 17.78 & -- & 3.8 & -- \\
&
& TD  & 5.78 & 3.08 & 0.068 & 1.8\% \\
\midrule
\multirow{2}{*}{$450\times360\times180$}
& \multirow{2}{*}{29.46M}
& FEM & 49.50 & -- & 13 & -- \\
&
& TD  & 10.94 & 4.52 & 0.12 & 0.95\% \\
\midrule
\multirow{2}{*}{$600\times480\times240$}
& \multirow{2}{*}{69.7M}
& FEM & 167.75 & -- & 30 & -- \\
&
& TD  & 20.08 & {8.35} & 0.18 & {0.61\%} \\
\bottomrule
\end{tabular}
\label{tab:4.1_cpu_performance}
\end{table}

\subsection{Large-scale simulations on GPU}
\label{sec:gpu_large_scale}

We next evaluate the TD method for larger scale simulations with the full GPU implementation. The domain size is increased to \(192\,\upmu\mathrm{m}\times76.8\,\upmu\mathrm{m}\times38.4\,\upmu\mathrm{m}\), allowing a larger number of grains and substantially larger problem sizes to be considered. The same physical and kinetic parameters are employed as in \tablename~\ref{tab:4.1_physical_parameters}. 

 Similar to previously, the accuracy of the TD results is assessed by comparing the TD predicted grain evolution and grain volume statistics with the corresponding reference FEM results. A larger mesh size with $800\times320\times160$ elements (approximately 41 million DoFs) with  500 initial grains is considered. The simulations are performed for 10,000 time steps with a step size of $\Delta t=5\times10^{-5},\mathrm{s}$. We illustrate the grain structures predicted by the FEM and TD at three time steps, $t=0$, $25$, and $50$ {ms}. 
 
\begin{figure}[htbp]
  \centering
  \subcaptionbox{FEM, \,$t=0\,\mathrm{ms}$}[0.27\textwidth]{%
    \includegraphics[width=\linewidth]{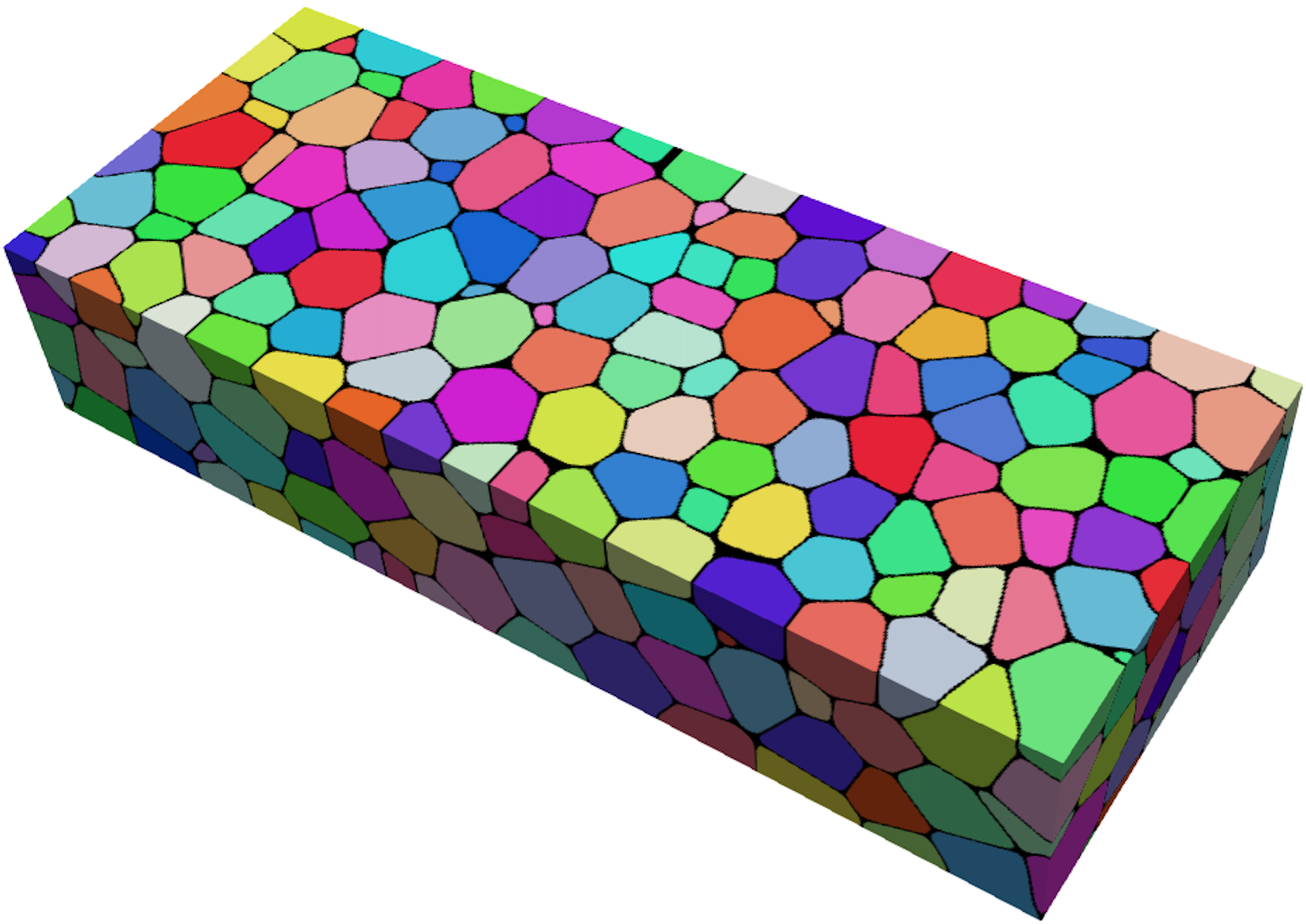}
  }
  \hspace{0.05\textwidth}
  \subcaptionbox{FEM, \,$t=25\,\mathrm{ms}$}[0.27\textwidth]{%
    \includegraphics[width=\linewidth]{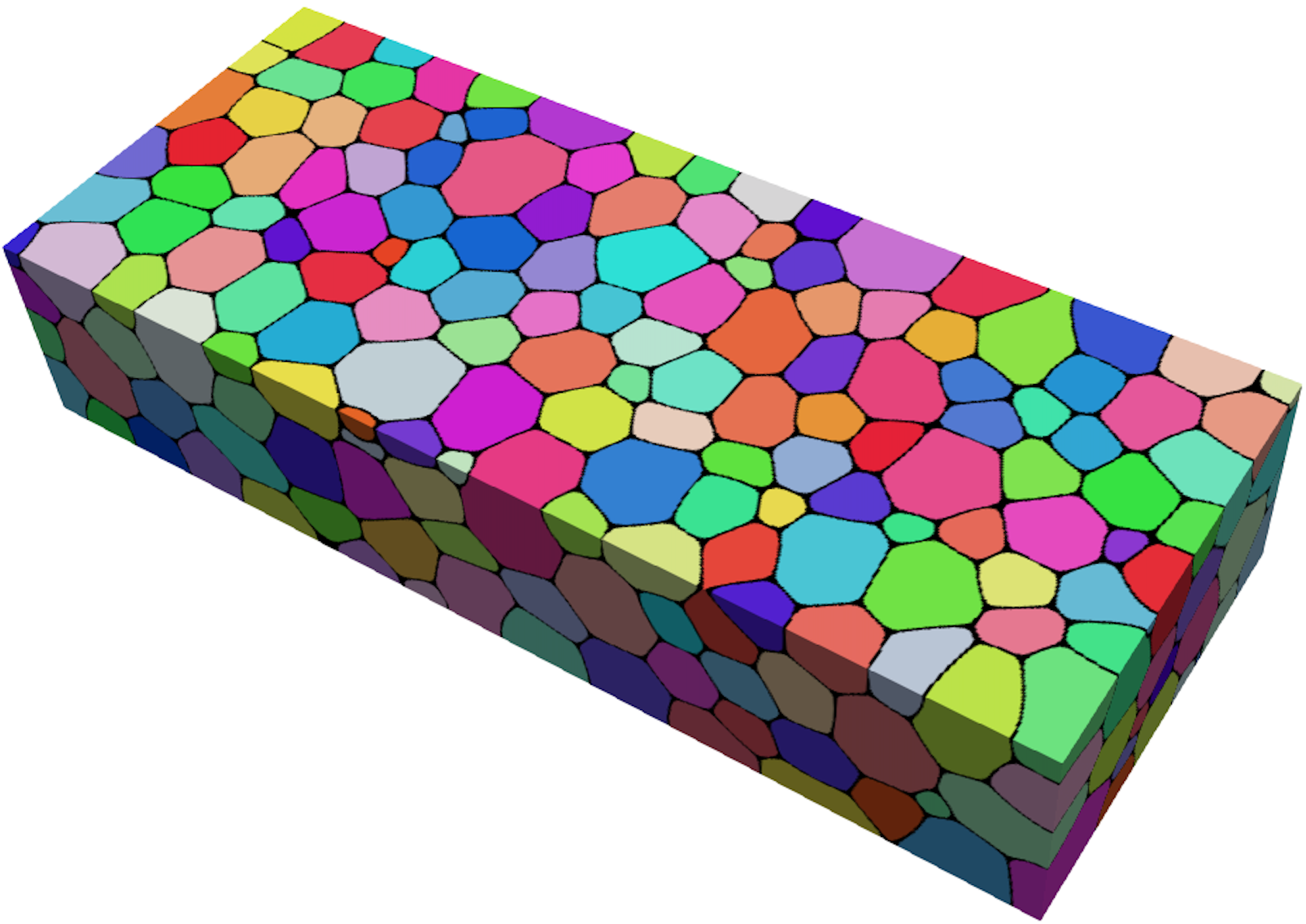}
  }
  \hspace{0.05\textwidth}
  \subcaptionbox{FEM, \,$t=50\,\mathrm{ms}$}[0.27\textwidth]{%
    \includegraphics[width=\linewidth]{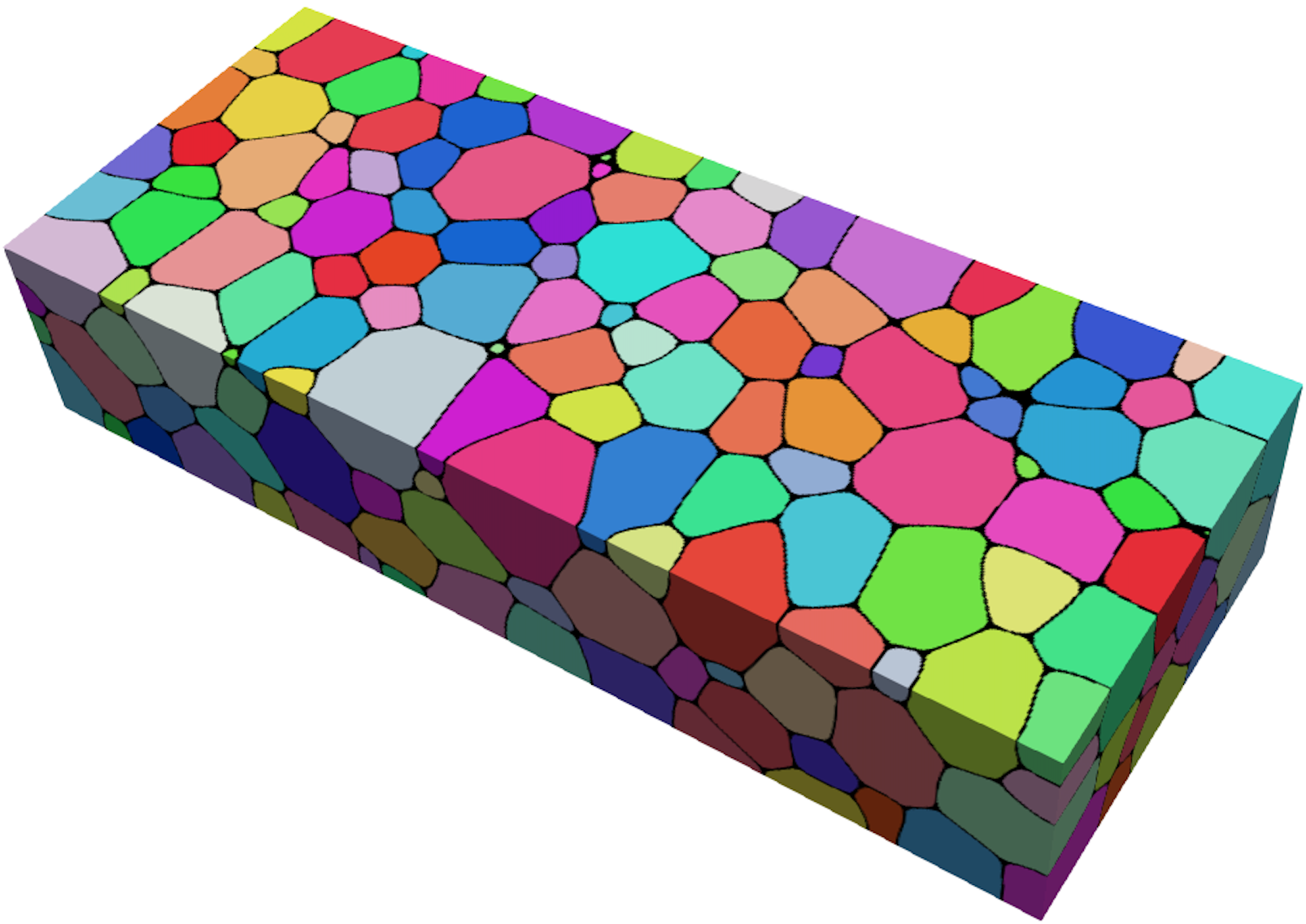}
  }

  \vspace{0.02\textwidth}

  \subcaptionbox{TD, \,$t=0\,\mathrm{ms}$}[0.27\textwidth]{%
    \includegraphics[width=\linewidth]{Figures/4.2/lt1.png}
  }
  \hspace{0.05\textwidth}
  \subcaptionbox{TD, \,$t=25\,\mathrm{ms}$}[0.27\textwidth]{%
    \includegraphics[width=\linewidth]{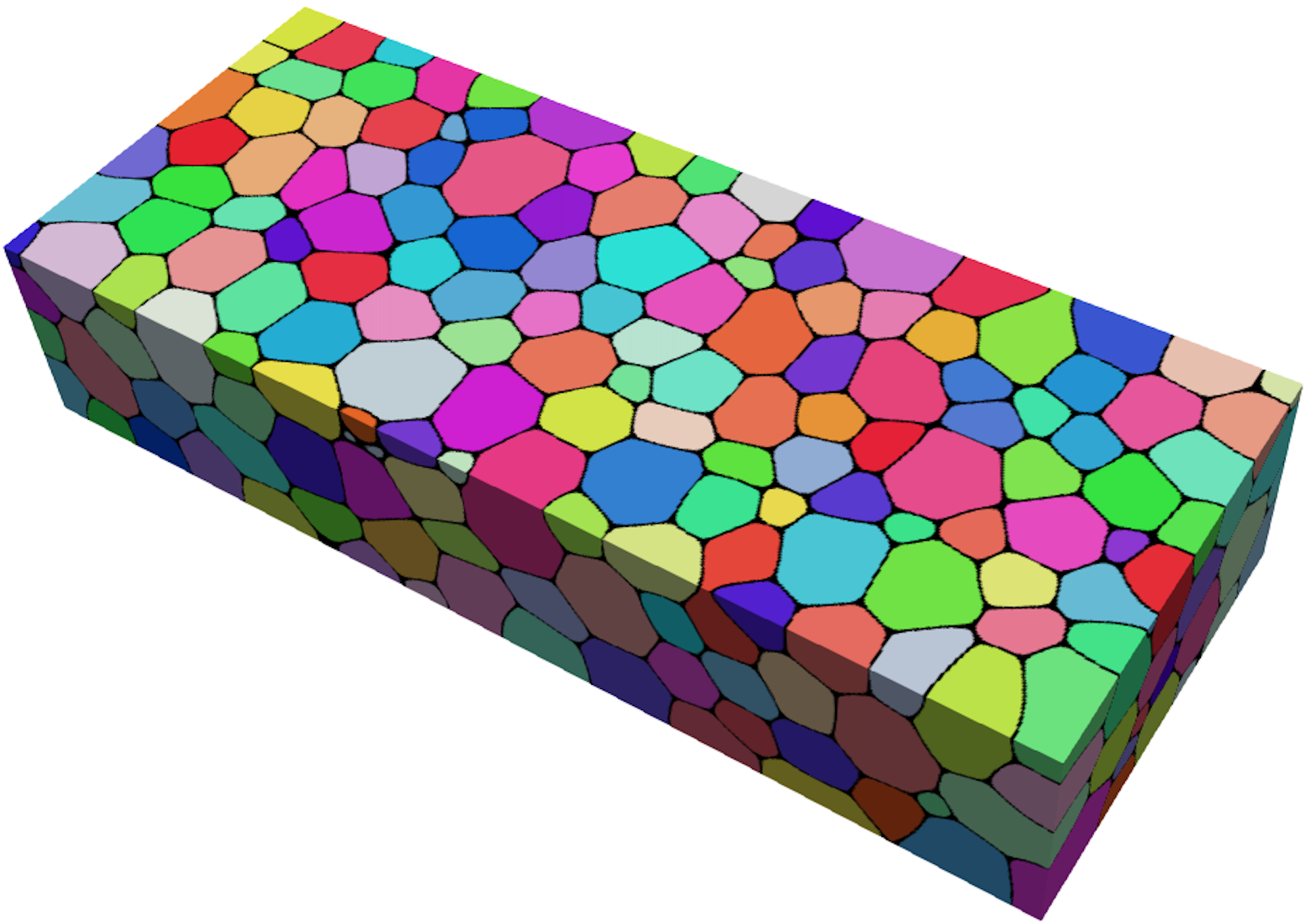}
  }
  \hspace{0.05\textwidth}
  \subcaptionbox{TD, \,$t=50\,\mathrm{ms}$}[0.27\textwidth]{%
    \includegraphics[width=\linewidth]{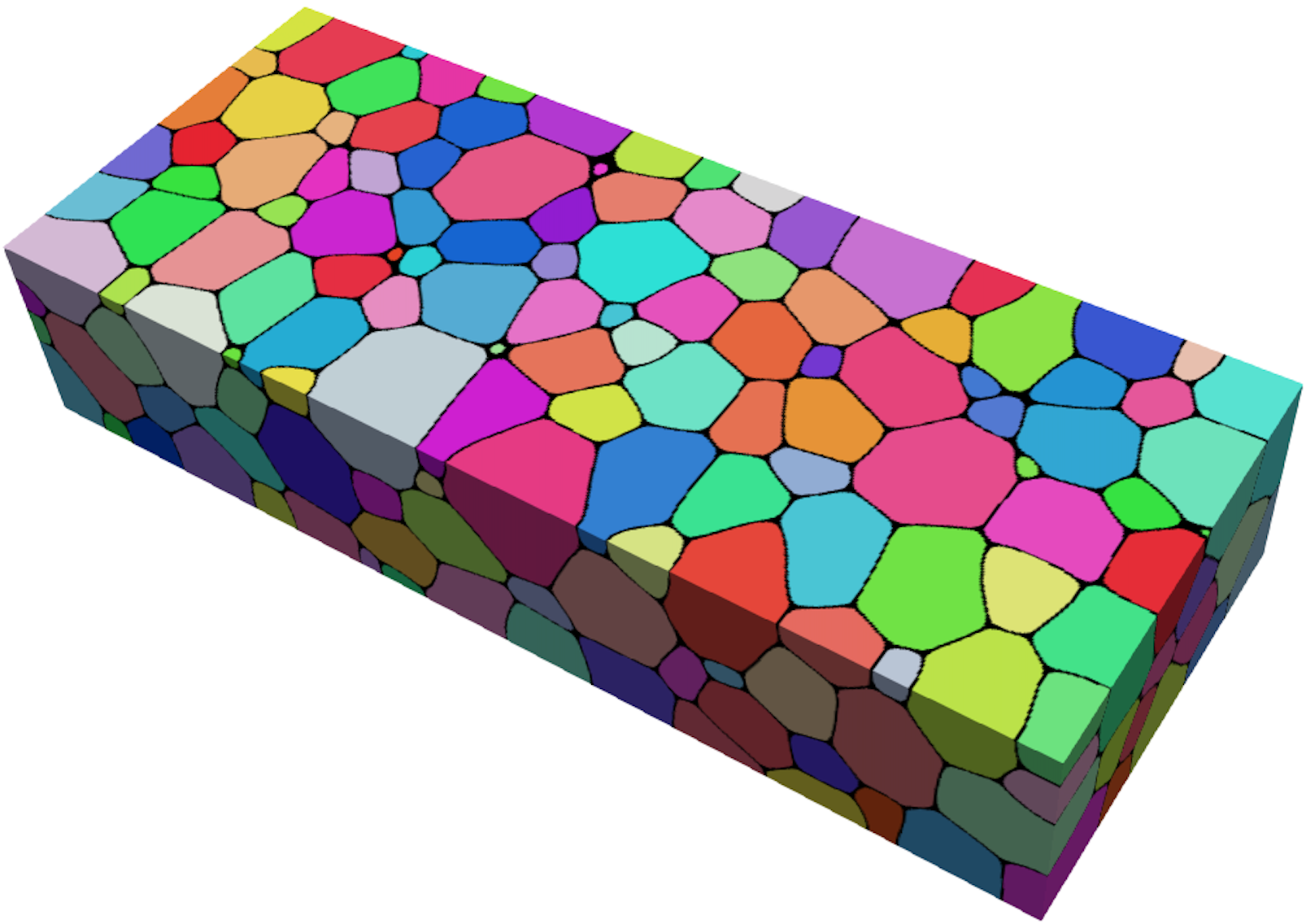}
  }

  \caption{Large-volume 3D grain growth obtained using GPU-based FEM and TD: (a)-(c) FEM and (d)-(f) TD.}
  \label{fig:4.2_gpu_grain_evolution}
\end{figure}

 As shown in Figure~\ref{fig:4.2_gpu_grain_evolution}, the larger computational domain and increased number of grains introduce substantially greater microstructural complexity, including  a more intricate grain boundary network, a wider variety of grain morphologies, and more extensive interactions/competitions among neighboring grains. These features present additional challenges for accurately resolving grain evolution across a large 3D domain, particularly as the local grain configurations and their spatial relationships continuously change during the simulation. Despite these challenges, the FEM and TD results exhibit close agreement in the overall microstructural networks and the spatial variations in individual grain morphologies. The TD results maintain consistent agreement with the reference FEM results as the microstructure evolves, without noticeable deterioration in the predicted grain configurations over the long-time evolution.

\begin{figure}[htbp]
  \centering
  \subcaptionbox{\,$t=0\,\mathrm{ms}$}[0.28\textwidth]{%
    \includegraphics[width=\linewidth]{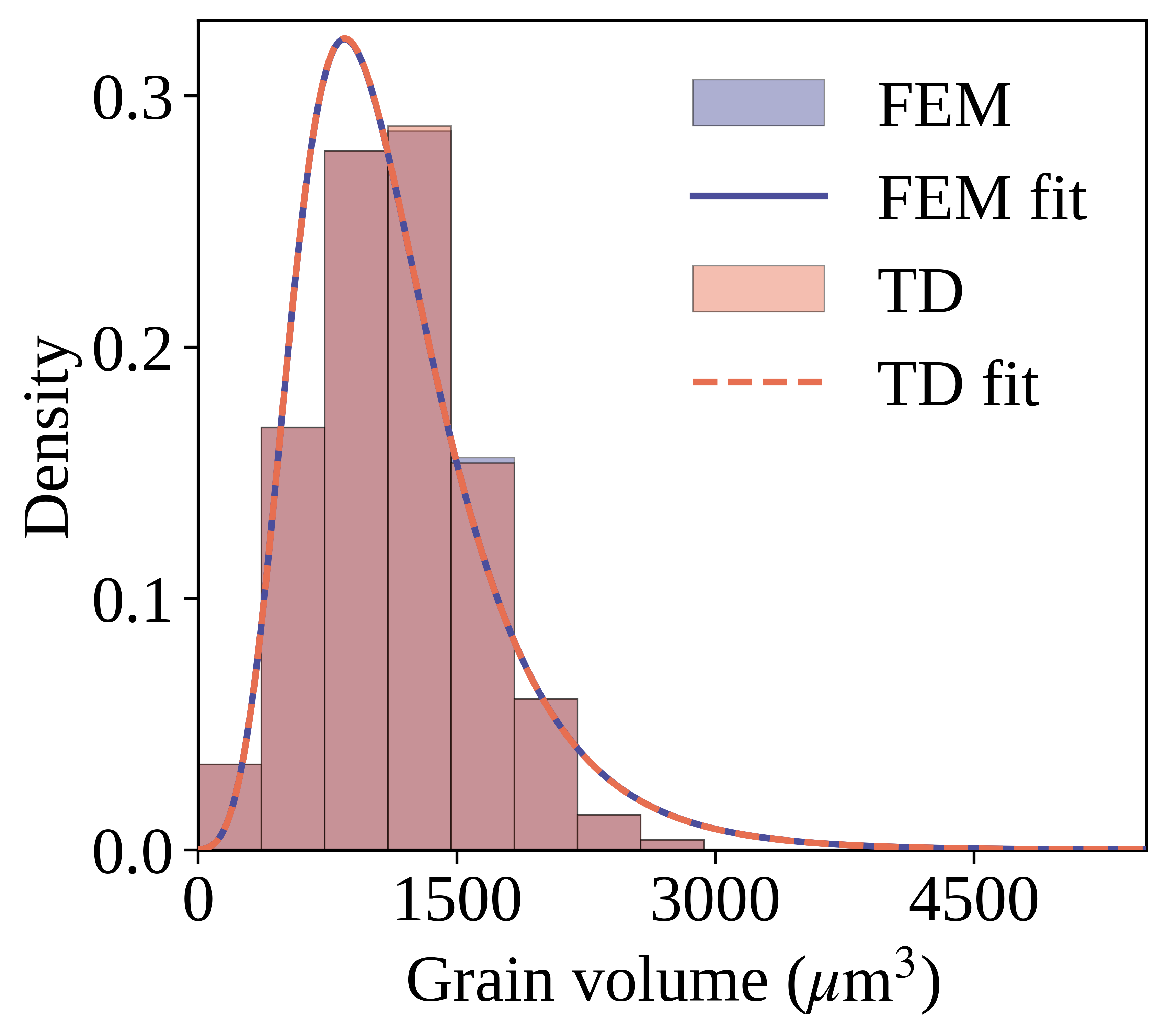}
  }
  \hspace{0.05\textwidth}
  \subcaptionbox{\,$t=25\,\mathrm{ms}$}[0.28\textwidth]{%
    \includegraphics[width=\linewidth]{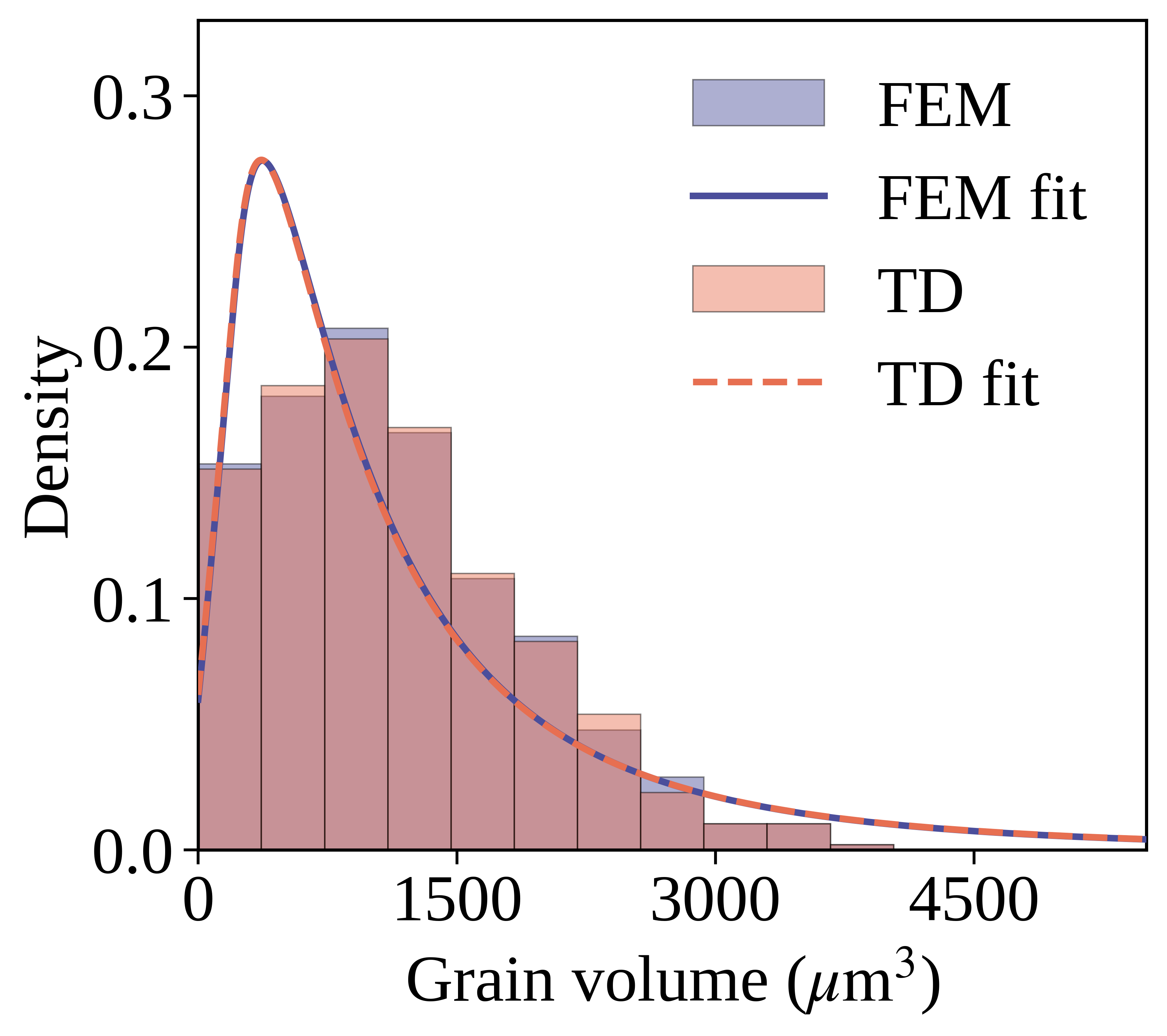}
  }
  \hspace{0.05\textwidth}
  \subcaptionbox{\,$t=50\,\mathrm{ms}$}[0.28\textwidth]{%
    \includegraphics[width=\linewidth]{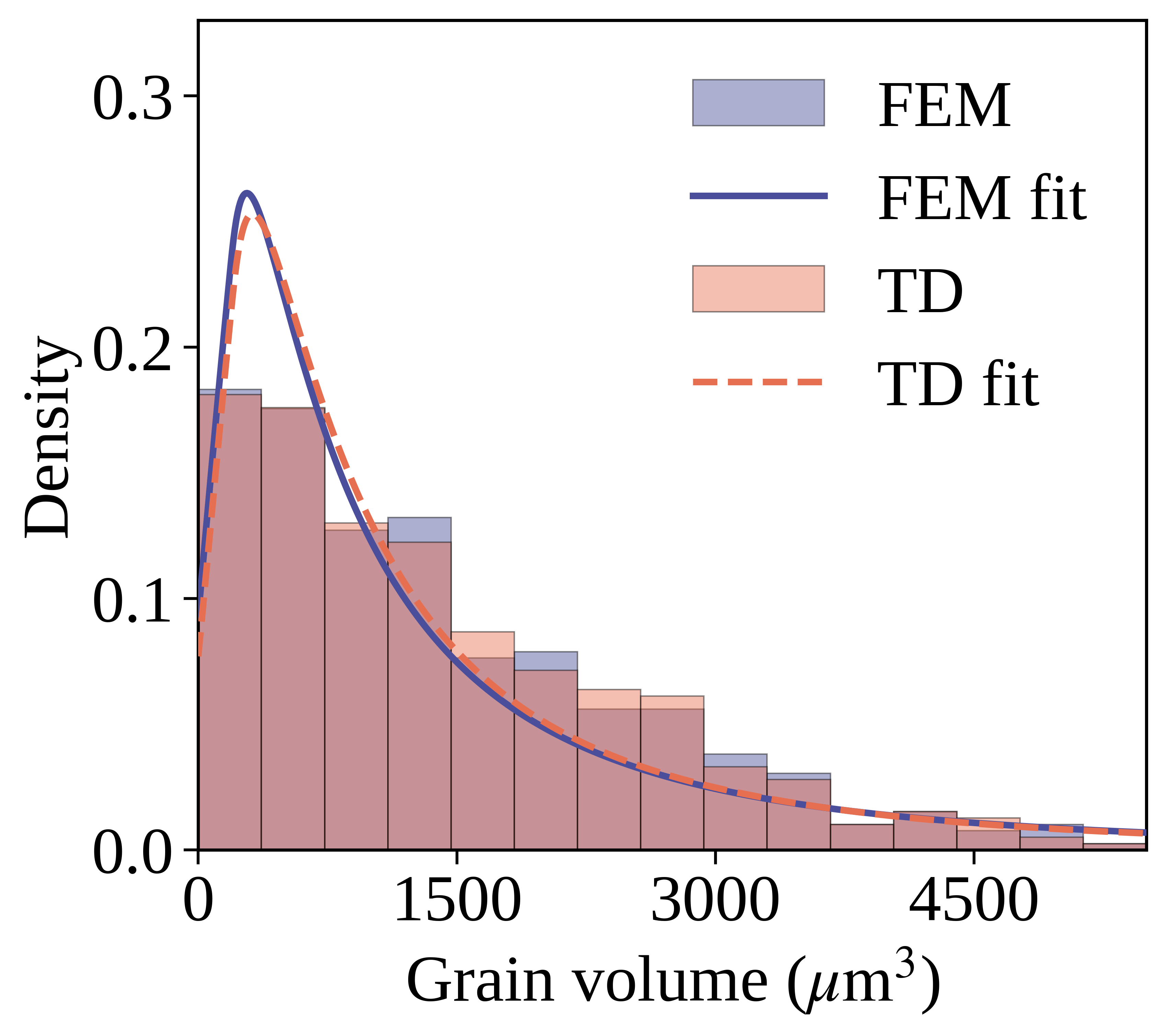}
  }

  \vspace{0.02\textwidth}

  \subcaptionbox{\,$t=0\,\mathrm{ms}$}[0.28\textwidth]{%
    \includegraphics[width=\linewidth]{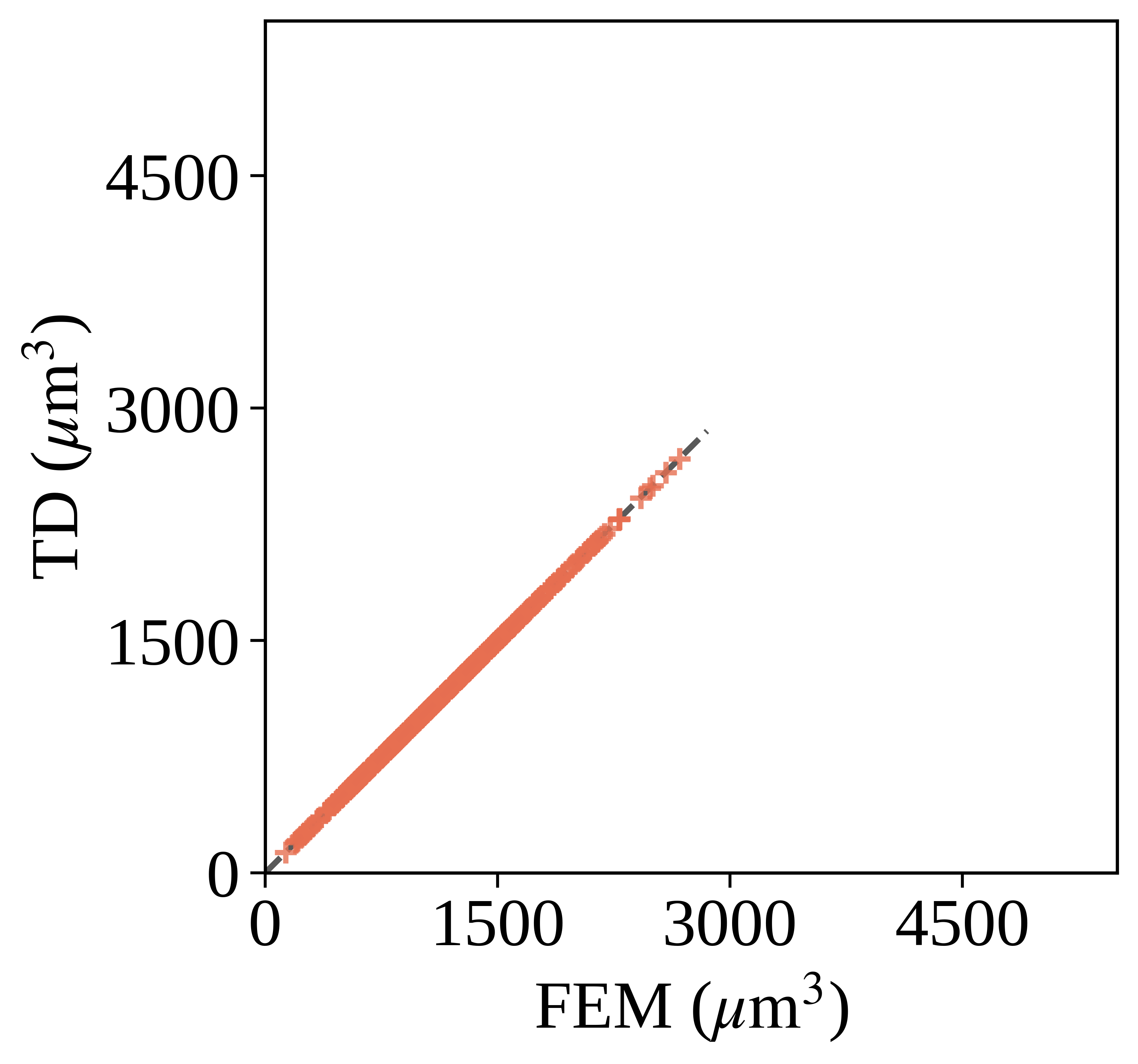}
  }
  \hspace{0.05\textwidth}
  \subcaptionbox{\,$t=25\,\mathrm{ms}$}[0.28\textwidth]{%
    \includegraphics[width=\linewidth]{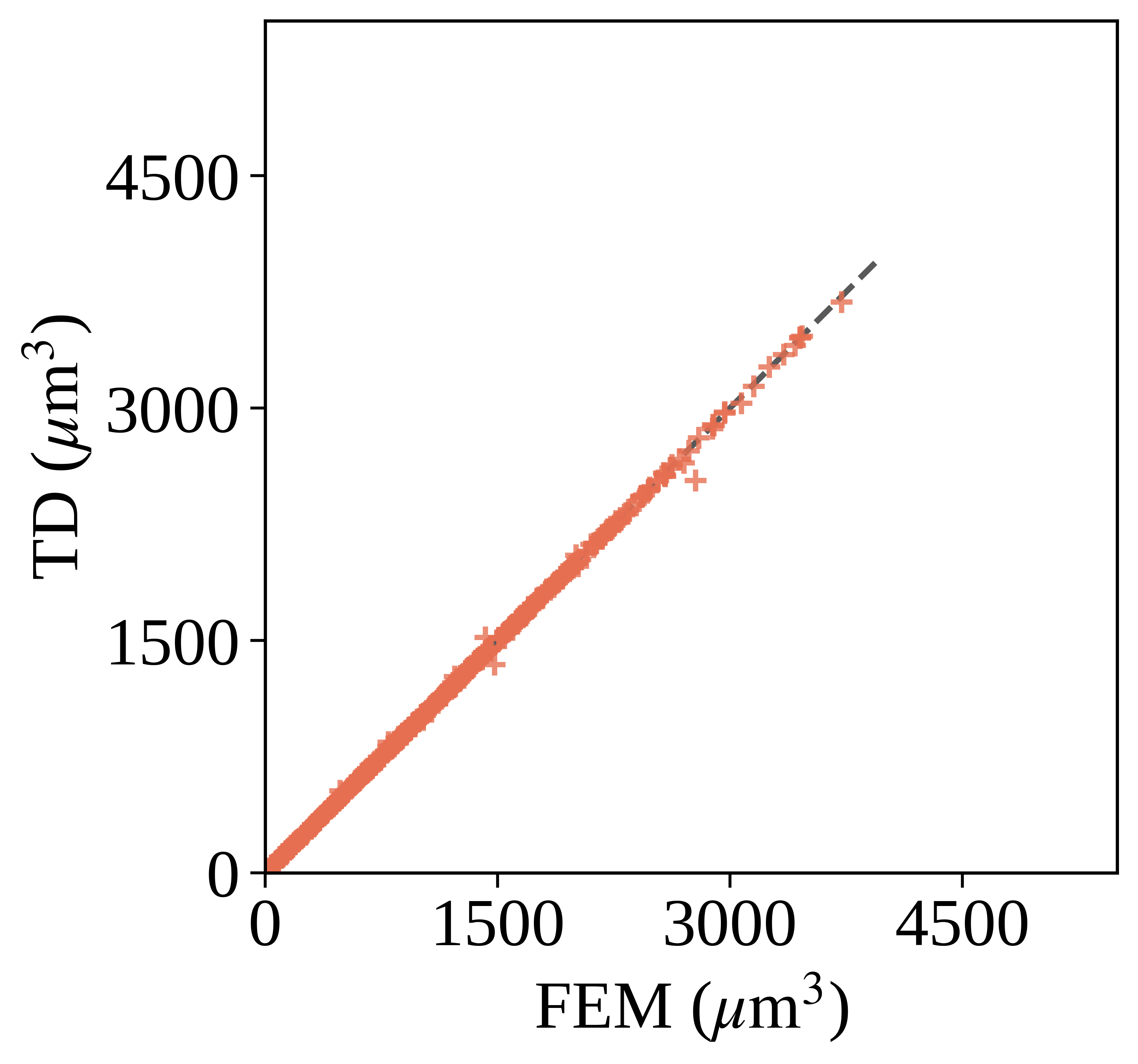}
  }
  \hspace{0.05\textwidth}
  \subcaptionbox{\,$t=50\,\mathrm{ms}$}[0.28\textwidth]{%
    \includegraphics[width=\linewidth]{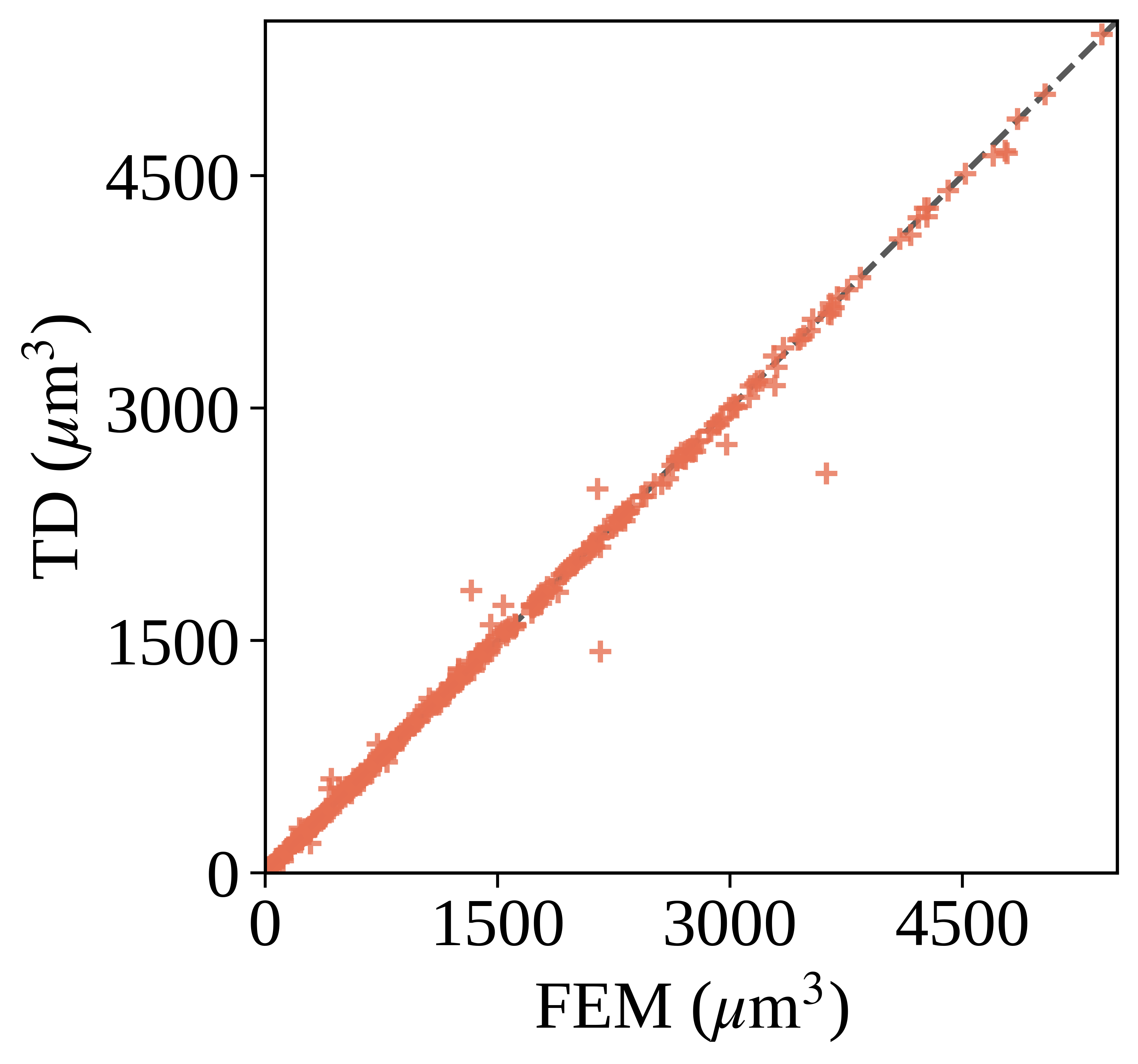}
  }

  \caption{Grain-volume comparison between GPU-based FEM and TD: (a)-(c) volume distributions with lognormal fits and (d)-(f) grain-wise volume comparison, with  MARE $=$ 0.047\%, 0.96\%, and 3.42\%, respectively.}
  \label{fig:4.2_gpu_grain_volume}
\end{figure}

Figure~\ref{fig:4.2_gpu_grain_volume} provides a quantitative assessment of the TD results by comparing individual grain volumes and their statistical distributions against the reference FEM results. The grain-volume distributions and corresponding lognormal fits exhibit close agreement at all three time steps, indicating that the TD method accurately captures the overall statistical characteristics of the evolving microstructure. The mean MAREs of individual grain volumes are 0.047\%, 0.96\%, and 3.42\% at $t=0$, $25$, and $50\,\mathrm{ms}$, respectively. Although the grain-wise error slightly increases as the simulation progresses, the TD solution continues to reproduce the overall grain volume distribution with good accuracy. This could be improved by further adjusting the convergence criterion and increasing the number of modes. These results confirm the capability of the TD method to handle increasingly complex grain configurations and accurately capture large-volume 3D microstructure evolution. 

To further evaluate the performance of  TD for large-scale simulations, we increase the mesh resolution to achieve billions of DoFs. Table~\ref{tab:4.2_gpu_performance} summarizes the computational performance and solution data size of the FEM and TD methods for three problem sizes. The TD method demonstrates substantial computational savings, with the speedup increasing from 1.92$\times$ at 41.40 million DoFs to 12.13$\times$ at 1.11 billion DoFs, corresponding to a runtime reduction of approximately 92\% for the largest case. Even more significant savings are achieved in data storage. For the billion-DoF simulation, the TD representation results in only 0.75~TB of data, compared with 160.55~TB for FEM, representing a remarkable 99.53\% reduction in storage requirements. These results highlight again the exceptional data efficiency of the TD method, together with its increasing computational advantages as the problem size grows, demonstrating its potential for large-scale phase-field simulations.

\begin{table}[h]
\centering
\caption{Performance of FEM and TD on  GPU.}
\small
\begin{tabular}{ccccccc}
\toprule
\textbf{Mesh Size} &
\textbf{Total DoFs} &
\textbf{Method} &
\textbf{Total Time (h)} &
\textbf{Speedup} &
\textbf{Data Size (TB)} &
\textbf{Relative Data Size} \\
\midrule
\multirow{2}{*}{$800\times320\times160$}
& \multirow{2}{*}{41.40M}
& FEM & 3.67 & -- & 6.62 & -- \\
&
& TD  & 1.91 & 1.92 & 0.25 & 3.7\% \\
\midrule
\multirow{2}{*}{$1600\times640\times320$}
& \multirow{2}{*}{329.42M}
& FEM & 19.90 & -- & 49.00 & -- \\
&
& TD  & 3.16 & 6.30 & 0.48 & 0.98\% \\
\midrule
\multirow{2}{*}{$2400\times960\times480$}
& \multirow{2}{*}{1.11B}
& FEM & 67.32 & -- & 160.55 & -- \\
&
& TD  & 5.55 & \textbf{12.13} & 0.75 & \textbf{0.47\%} \\
\bottomrule
\end{tabular}
\label{tab:4.2_gpu_performance}
\end{table}

\FloatBarrier
\subsection{Particle-mediated grain growth}
\label{sec:pinning}

As an application of the large-scale phase-field simulations enabled by the proposed TD method, we consider a grain growth problem involving dispersed incoherent second-phase particles, which requires fine mesh resolution to resolve grain-particle interactions. The problem can be modeled using a modified phase-field formulation \cite{Moelans2005}, in which the second-phase particles are represented by a prescribed spatial field $P(\boldsymbol{x})$ that remains fixed throughout the simulation. The homogeneous free energy in this case becomes
\begin{equation}
\label{eq:4.3_pinning_free_energy}
f(\{\phi\},P)
=
W\left[
\sum_{i=1}^{N}\left(\frac{\phi_i^4}{4}-\frac{\phi_i^2}{2}\right)
+\gamma_{PF}\sum_{i=1}^{N}\sum_{j>i}^{N}\phi_i^2\phi_j^2
+\frac{1}{4}
\right]
+\epsilon_{\mathrm{pin}}P(\boldsymbol{x})\sum_{i=1}^{N}\phi_i^2,
\end{equation}
where $\epsilon_{\mathrm{pin}}>0$ is a parameter that controls the strength of grain-particle interactions, and the particle field is defined as
\begin{equation}
P(\boldsymbol{x})=
\begin{cases}
1, & \boldsymbol{x}\ \text{inside a second-phase particle},\\
0, & \text{otherwise}.
\end{cases}
\label{eq:4.3_particle_field}
\end{equation}

The additional term, $\epsilon_{\mathrm{pin}}P(\boldsymbol{x})\sum_{i=1}^{N}\phi_i^2$, introduces an energetic penalty for nonzero OPs within the second-phase particle regions. Consequently, it drives the OPs toward zero within these regions, effectively preventing the matrix grains from occupying the particles. Since the particle field $P(\boldsymbol{x})$ remains fixed throughout the simulation, the particles act as stationary obstacles to grain boundary migration. As grain boundaries encounter these particles, their motion is impeded, giving rise to the particle-pinning effect (Zener pinning) \cite{Moelans2005}, which suppresses grain growth and influences the resulting microstructure evolution.

In the present simulation, we choose $\epsilon_{\mathrm{pin}}=1.5\,W$. The particle volume fraction is set to 0.04, and the particle radii are sampled from a lognormal distribution with a mean of $0.6\,\upmu\mathrm{m}$ and a coefficient of variation 0.10. The computational domain is $192\,\upmu\mathrm{m}\times76.8\,\upmu\mathrm{m}\times38.4\,\upmu\mathrm{m}$ and is discretized using a mesh containing $1600\times640\times320$ elements, corresponding to a uniform grid spacing of $0.12\,\upmu\mathrm{m}$, which is one-fifth of the mean particle radius. This resolution satisfies the recommended minimum requirement for adequately resolving the second-phase particles \cite{Moelans2005}.
The fine mesh used here is therefore necessary not only for resolving the grain boundaries, but also for adequately capturing the interaction between the dispersed particles and the evolving grains.

\begin{figure}[htbp]
  \centering
  \subcaptionbox{w/o particles, \,$t=0\,\mathrm{ms}$}[0.27\textwidth]{%
    \includegraphics[width=\linewidth]{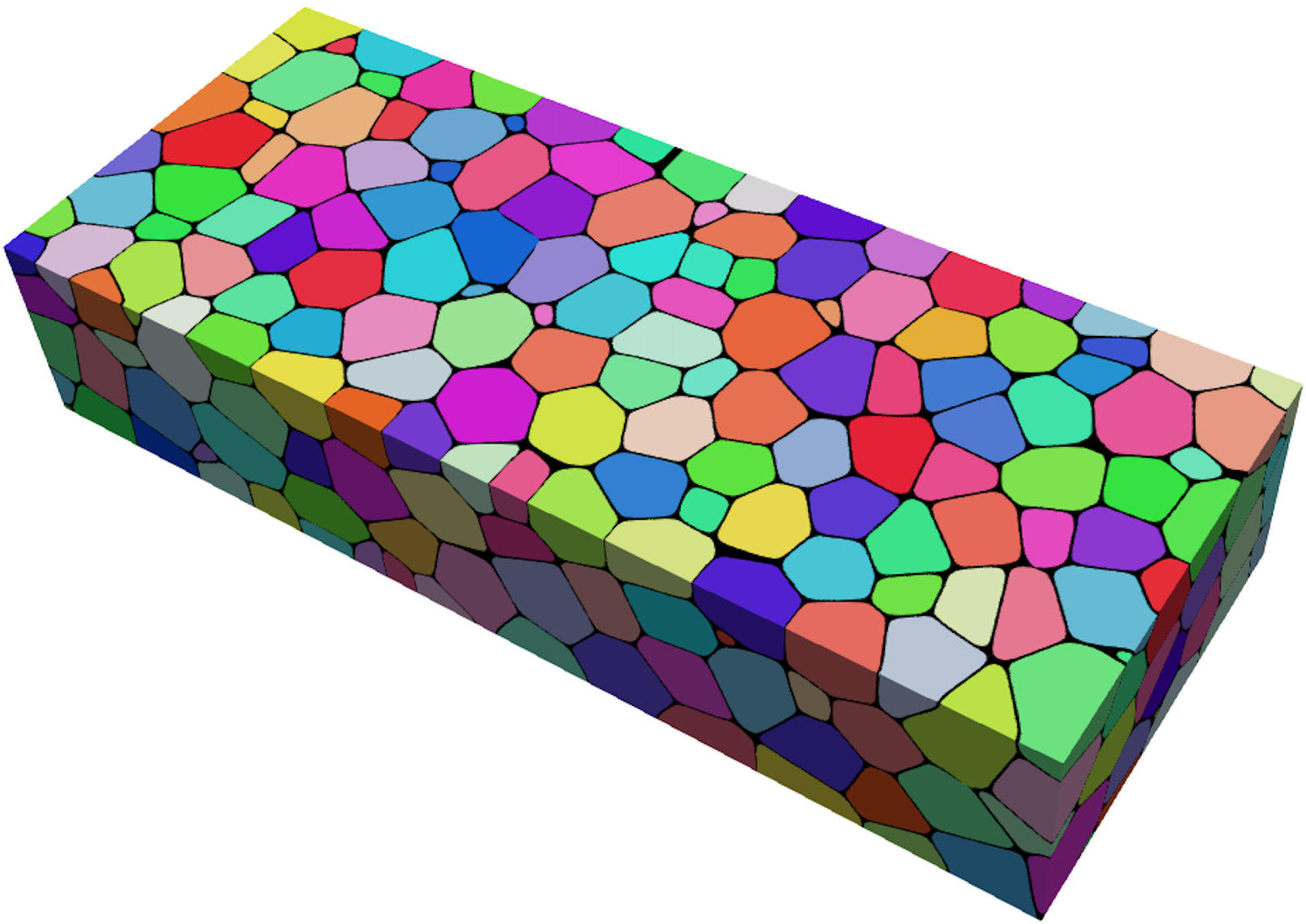}
  }
  \hspace{0.05\textwidth}
  \subcaptionbox{w/o  particles, \,$t=25\,\mathrm{ms}$}[0.27\textwidth]{%
    \includegraphics[width=\linewidth]{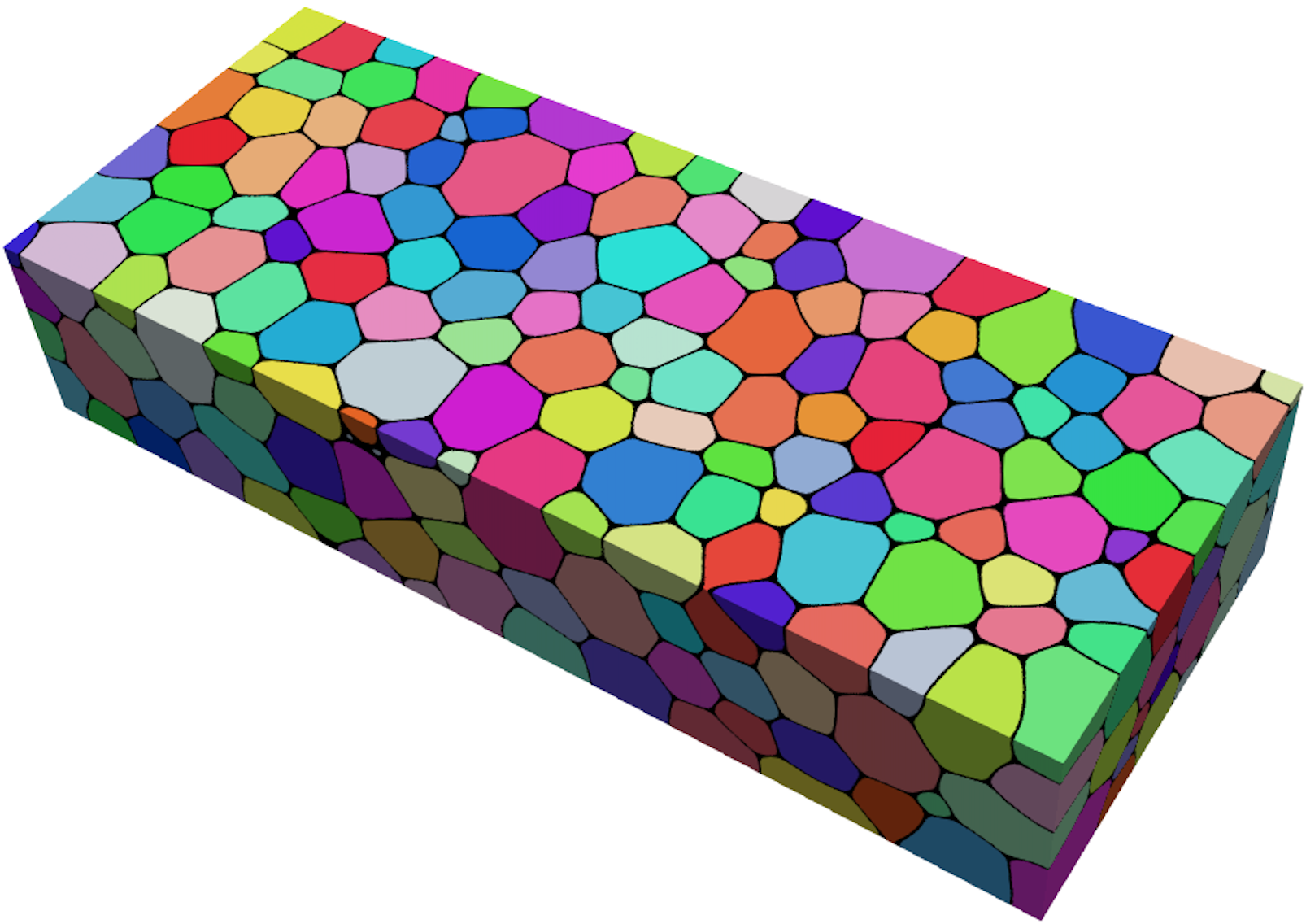}
  }
  \hspace{0.05\textwidth}
  \subcaptionbox{w/o  particles, \,$t=50\,\mathrm{ms}$}[0.27\textwidth]{%
    \includegraphics[width=\linewidth]{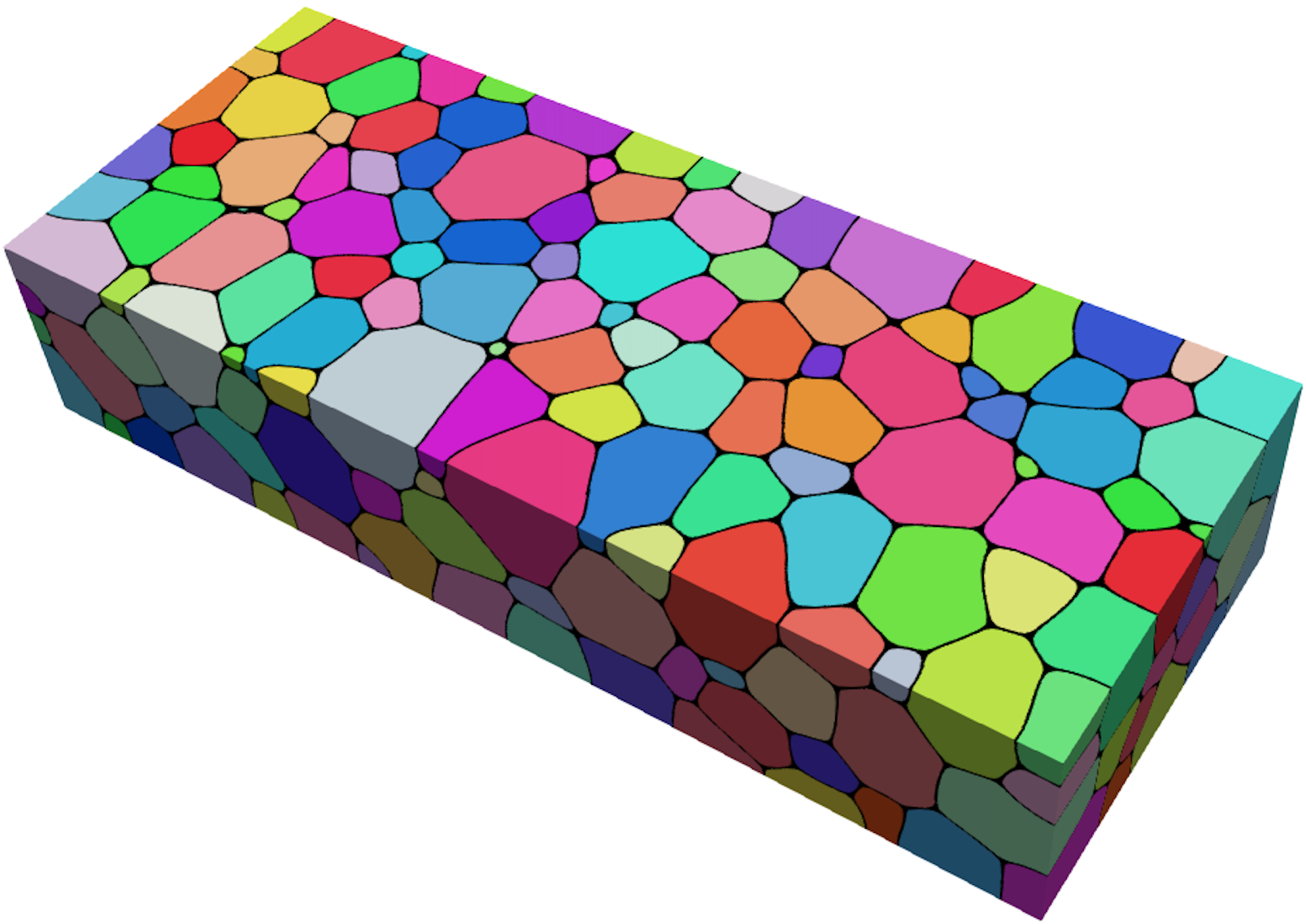}
  }

  \vspace{0.02\textwidth}

  \subcaptionbox{w/ particles,\,$t=0\,\mathrm{ms}$}[0.27\textwidth]{%
    \includegraphics[width=\linewidth]{Figures/4.3/pin0.png}
  }
  \hspace{0.05\textwidth}
  \subcaptionbox{w/ particles,\,$t=25\,\mathrm{ms}$}[0.27\textwidth]{%
    \includegraphics[width=\linewidth]{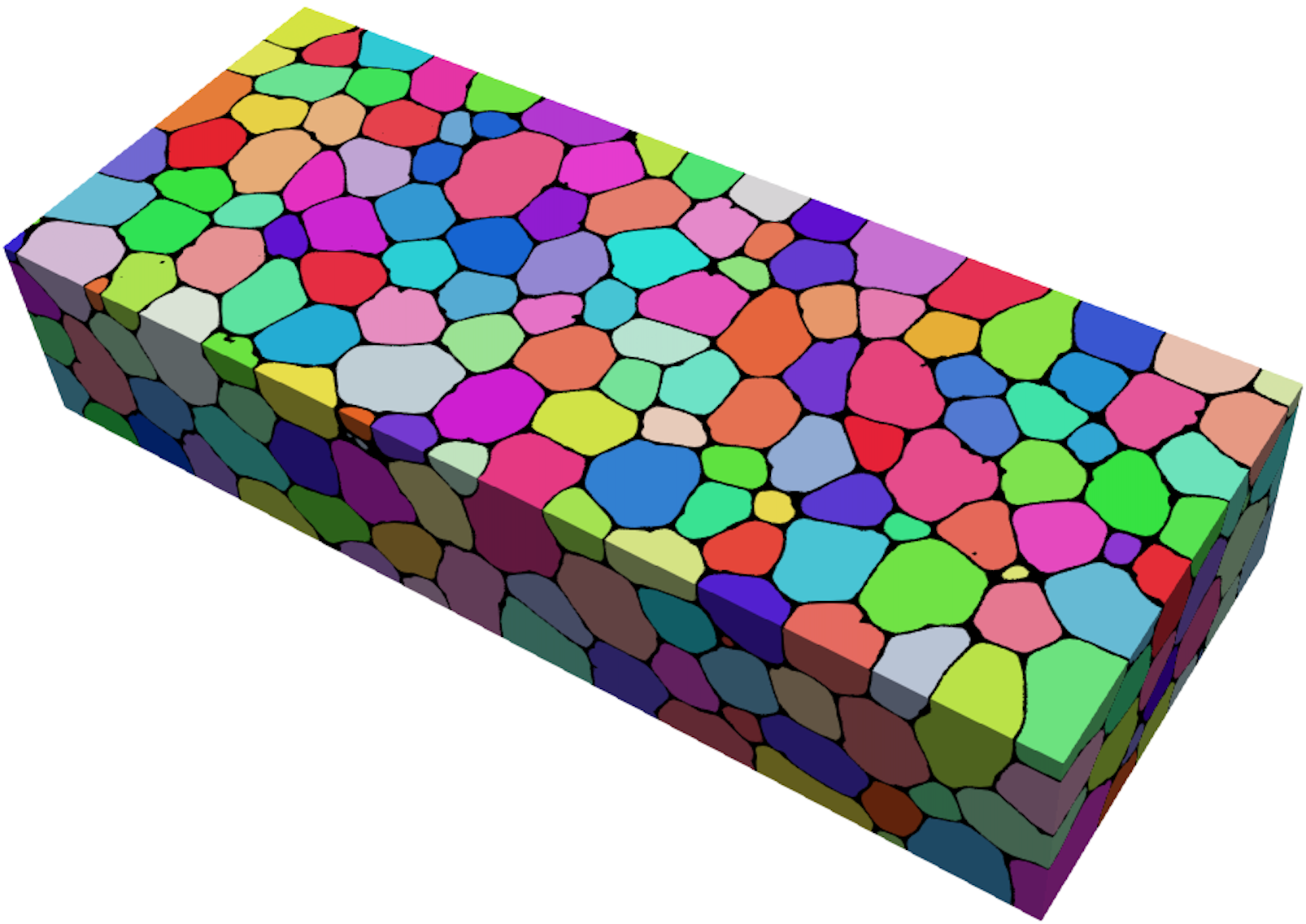}
  }
  \hspace{0.05\textwidth}
  \subcaptionbox{w/ particles,\,$t=50\,\mathrm{ms}$}[0.27\textwidth]{%
    \includegraphics[width=\linewidth]{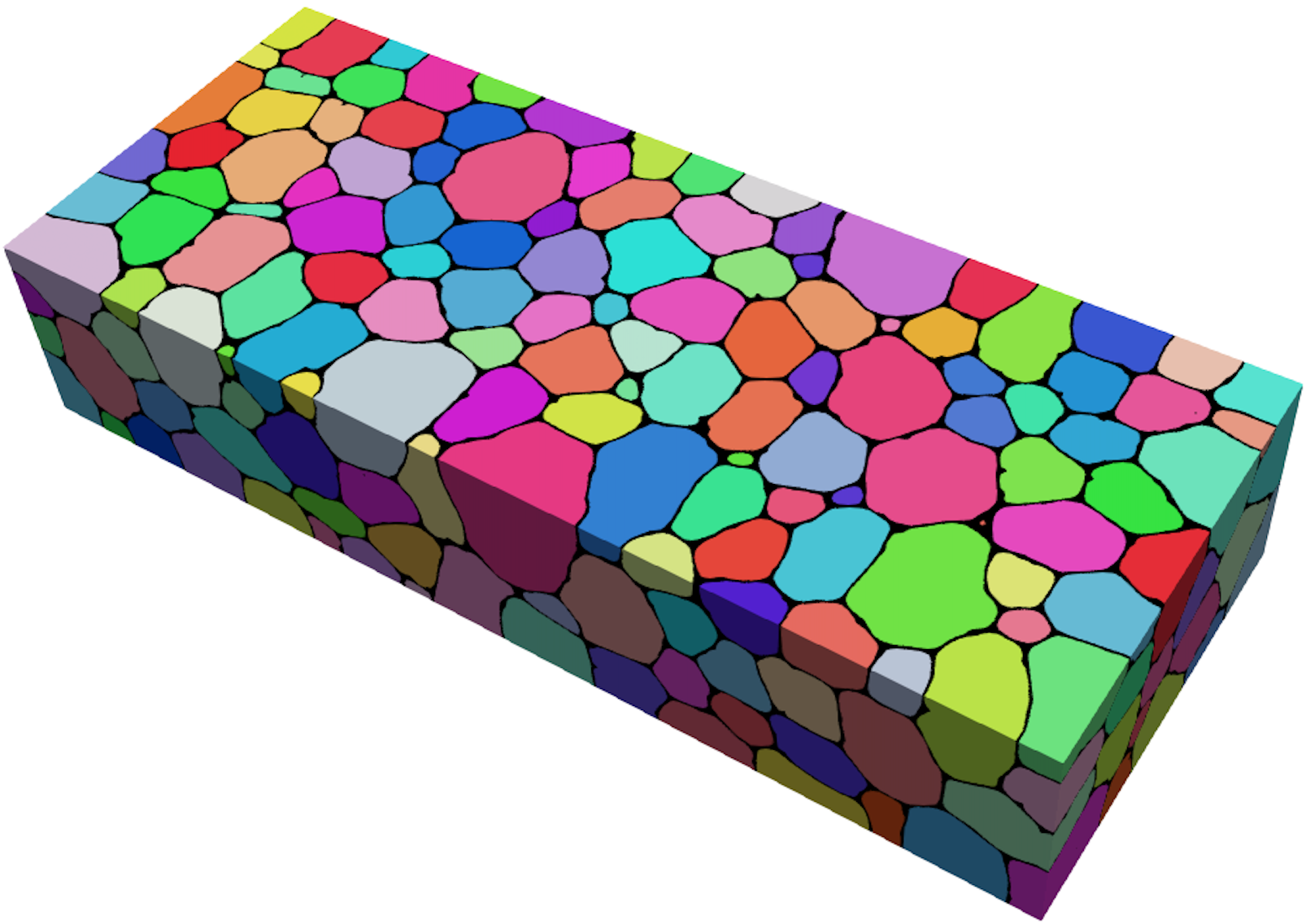}
  }

  \caption{Grain growth with/without second-phase particles: (a)-(c) without particles, (d)-(f) with particles.}
  \label{fig:4.3_pin_grain_evolution}
\end{figure}

The results are illustrated in Figure~\ref{fig:4.3_pin_grain_evolution}. We compare the simulated grain structures with and without the second-phase particles  at three time steps $t=0$, $25$, and $50\,\mathrm{ms}$. Both cases start from the same initial grain structure. As time proceeds, more pronounced grain coarsening is observed in the case without the particles, with larger grains growing while smaller grains gradually disappear. In contrast, the presence of second-phase particles suppresses grain growth through the particle-pinning effect, resulting in a finer grain structure, as expected. The difference becomes increasingly evident at $t=25$ and $50\,\mathrm{ms}$, demonstrating the significant influence of the second-phase particles on the overall grain-coarsening process.

\begin{figure}[htbp]
  \centering
    \includegraphics[width=0.9\linewidth]{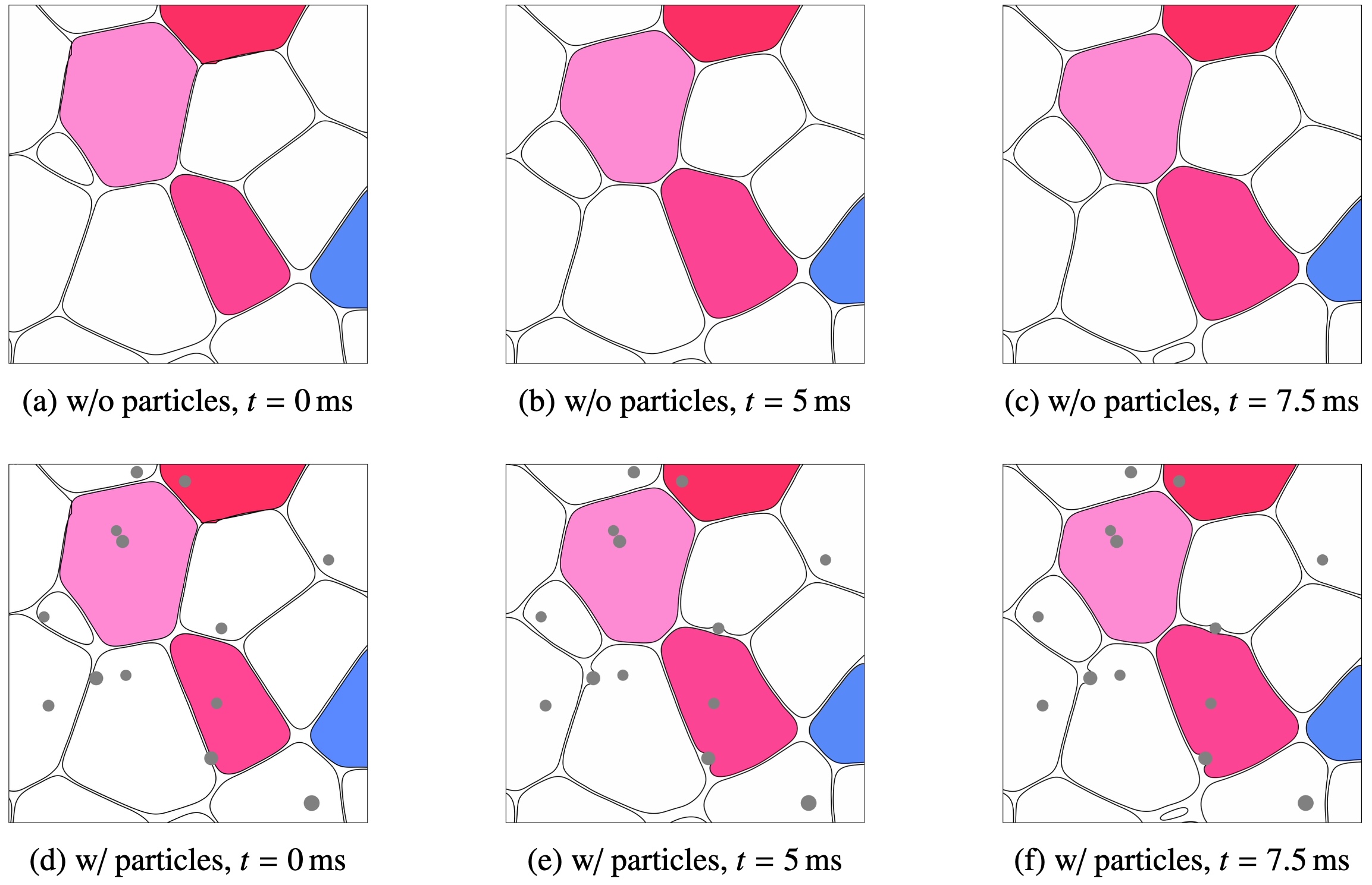}
  \caption{Top surface structures with/without particles: (a)-(c) without particles, (d)-(f) with particles (gray dots).}
  \label{fig:4.3_zener_pinning}
\end{figure}

Figure~\ref{fig:4.3_zener_pinning} shows a local region on the top surface to illustrate the detailed particle-pinning effect. In the case without particles, the grain boundaries migrate continuously during grain growth. In the case with particles, the local motion of the grain boundaries is impeded when they encounter the particles, while the boundary segments between neighboring particles continue to migrate. This produces a modified local grain morphology and reduces the overall grain boundary mobility, which is consistent with the characteristic Zener pinning mechanism and the work reported by \cite{Moelans2005}. This demonstrates again the computational capability of the TD framework for large-scale simulations and its flexibility to accommodate different phase-field models for complex microstructure evolution.

\section{Conclusion}

The TD based model reduction framework for large-scale multiphase-field simulations of microstructure evolution has been developed. The proposed framework combines spatially decomposed low-rank approximation, subspace mode updating, and GPU/CPU parallel computing. The framework has been applied to simulate complex grain growth in polycrystalline materials. The key findings are summarized as follows.

\begin{enumerate}
    \item The TD method can accurately simulate the grain growth behavior obtained from the full-order FE simulations, including grain morphology, grain volume, and the overall coarsening characteristics.
    \item The subspace mode-updating algorithm can effectively control the number of modes and improve the compactness of the TD representation, providing a practical balance between computational cost and data reduction.
    \item The TD method demonstrates increasing computational and data advantages as the problem size grows, confirming its efficiency for large-scale microstructure simulations.
    \item The GPU implementation further improves the computational efficiency by exploiting grain-level parallelism, compact GPU-resident storage, and batched one-dimensional solves. For the largest case considered, containing more than one billion DoFs, the TD method achieves more than $10\times$ speedups while resulting in less than 1\% of the solution data over the GPU-based FEM.
    \item The proposed framework can readily accommodate various phase-field formulations. Simulations with dispersed incoherent second-phase particles reproduce the characteristic suppression of grain coarsening and local Zener-pinning behavior reported in the literature.
\end{enumerate}

Our future work will extend the framework to multi-GPU and distributed-memory computing for even larger simulations. Another important direction is the application to advanced manufacturing processes, enabling efficient large-scale simulations of microstructure evolution and the investigation of process-microstructure linkage.


\section*{Acknowledgements}
CY and YL would like to acknowledge the support of University of Maryland Baltimore County through the startup fund. {This work was carried out using the High Performance Computing Facility (HPCF) at University of Maryland Baltimore County.}

\bibliographystyle{main.bst} 
\bibliography{main.bib}






\end{document}